\pdfoutput=1
\newcommand*{\ATLASLATEXPATH}{latex/}
\documentclass[UKenglish,cernpreprint,texlive=2016]{\ATLASLATEXPATH atlasdoc}

\usepackage{\ATLASLATEXPATH atlaspackage}
\usepackage{\ATLASLATEXPATH atlasbiblatex}
\usepackage{placeins}
\usepackage{\ATLASLATEXPATH atlascontribute}

\usepackage{\ATLASLATEXPATH atlasphysics}

\usepackage{longtable}
\usepackage{cancel}
\usepackage{graphicx}
\usepackage{caption}
\usepackage{subfig}
\usepackage{makecell}
\usepackage{pbox}

\graphicspath{{logos/}{figures/}}

\usepackage{HIGG-2016-09-PAPER-defs}

\AtlasTitle{Search for doubly charged scalar bosons decaying into same-sign $W$ boson pairs with the ATLAS detector}
\AtlasAuthorContributor{The ATLAS Collaboration}{}{}

\AtlasRefCode{HIGG-2016-09}
\AtlasNote{HIGG-2016-09}

\PreprintIdNumber{CERN-EP-2018-188}

 \AtlasJournal{Eur.\ Phys.\ J.\ C}
\AtlasJournalRef{\EPJC 79 (2019) 58}
\AtlasDOI{10.1140/epjc/s10052-018-6500-y}

\AtlasAbstract{%
A search for doubly charged scalar bosons decaying into $W$ boson pairs is presented. It uses a data sample from proton--proton collisions corresponding to an integrated luminosity of $36.1$~fb$^\mathrm{-1}$ collected by the ATLAS detector at the LHC at a centre-of-mass energy of 13~\TeV in 2015 and 2016.  This search is guided by a model that includes an extension of the Higgs sector through a scalar triplet, leading to a rich phenomenology that includes doubly charged scalar bosons  $H^{\pm\pm}$. Those bosons are produced in pairs in proton--proton collisions and decay predominantly into electroweak gauge bosons $H^{\pm\pm}\rightarrow W^{\pm} W^{\pm}$. Experimental signatures with several leptons, missing transverse energy and jets are explored. No significant deviations from  the Standard Model predictions are found. The parameter space of the benchmark model is excluded at 95\% confidence level for $H^{\pm\pm}$ bosons with masses between 200 and 220 \GeV.}

\begin{document}

\maketitle
\clearpage
\section{Introduction}
\label{sec:intro}
An extension of the scalar sector of the Standard Model (SM) is possible in the context of type II seesaw models~\cite{Schechter:1980gr}, originally conceived to explain the smallness of the neutrino masses.
In the model investigated in this paper, the scalar sector includes 
a hypercharge $Y=2$ scalar triplet, $\Delta$, in addition to the SM scalar doublet $H$~\cite{Perez:2008ha, Arhrib:2011uy}. 
Electroweak symmetry breaking (EWSB) is achieved if the neutral components of $H$ and $\Delta$ acquire vacuum expectation values, $v_{d}$ and $v_{t}$ respectively. After the EWSB, the mixing between these fields results in seven scalar bosons: $H^{\pm\pm}$, $H^{\pm}$, $A^{0}$ (CP odd), $H^{0}$ (CP even), $h^{0}$ (CP even). A small mixing between the CP-even scalars 
allows $h^{0}$ to have the expected properties of the SM Higgs boson. In addition, the triplet-neutrino Yukawa term 
 provides non-zero neutrino masses proportional to the vacuum expectation value of the triplet $v_{t}$.  
 Constraints from electroweak precision measurements lead to an upper bound on $\vt$ of around $1$~\GeV. This range is significantly lower than the electroweak scale and matches the need for small values suggested by the natural association of $\vt$ with the neutrino masses. 

The assumption of a non-zero $\vt $, of the order of a hundred \MeV, opens the possibility for the doubly charged boson to decay into a pair of same-sign $W$ bosons, \hppd, while the leptonic decays \hppl are suppressed with increasing $\vt$~\cite{Kanemura:2014goa, kang:2014jia}. 
Extensive searches for leptonic decays \hppl have been performed at various colliders~\cite{Abbiendi:2001cr,Aktas:2006nu,Aaltonen:2011rta, EXOT-2012-18,CMS-HIG-12-005,EXOT-2016-07}, where \hpp{} bosons with masses up to about 800~\GeV have been excluded. 
Moreover, searches for \hppd decays have been performed by the CMS Collaboration in the context of single $H^{\pm\pm}$ production through vector-boson fusion at large $\vt$ (of order of tens of~\GeV)~\cite{Khachatryan:2014sta, Sirunyan:2017ret} for a model with two Higgs triplets~\cite{Georgi:1985nv}. For that model, a custodial symmetry avoids large contributions to the electroweak precision observables~\cite{Englert:2013zpa}. In contrast, the \hppd decay mode has not been directly searched for so far for small values of $\vt$, where the vector-boson fusion is suppressed.

The present paper focuses on the phenomenology of doubly charged scalar bosons $H^{\pm\pm}$ that can be produced in pairs at the Large Hadron Collider (LHC) and decay into $W$ bosons. 
The triplet vacuum expectation value is taken to be $\vt=0.1$~\GeV such that only the \hppd decays are relevant, leading to final states with four $W$ bosons. The mixing  between the CP-even scalars is taken to be $10^{-4}$ and the remaining five Yukawa parameters in the potential are adjusted to obtain a given  $H^{\pm\pm}$ mass hypothesis while requiring $h^{0}$ to have a mass of $125$~\GeV. 
The corresponding cross-section calculation is performed for on-shell $W$ bosons, and therefore only the region $m_{H^{\pm\pm}} > 200$~\GeV is considered in the present analysis. 

The four-boson final states are identified by the presence of light charged leptons (electrons or muons), missing transverse momentum, and jets.
The analysis uses three final states defined according to the number of light leptons: same-sign (SS) dilepton channel (\ll), trilepton channel (\lll) and four-lepton channel (\llll). 
Similar final states were used for other searches for new phenomena in ATLAS~\cite{EXOT-2016-22,Aaboud:2018jiw,SUSY-2016-14}. However, the previously  searched signal topologies differ significantly from  those targeted in the present analysis and a dedicated event selection optimisation is therefore applied.  

This paper includes a description of the experimental set-up in Section~\ref{sec:detector}, followed by a description of  
the simulation used in the analysis in Section~\ref{sec:Simulation}.
The event selection and background estimations for the three explored signatures
are described in Section \ref{sec:EventSelection}. The signal region optimisation is described in Section~\ref{sec:SignalRegion}. The systematic uncertainties are presented in Section~\ref{sec:Systematics}. The results are shown in Section~\ref{sec:Results}, followed by the conclusions in Section~\ref{sec:Conclusions}.

\section{ATLAS detector}
\label{sec:detector}
\newcommand{\AtlasCoordFootnote}{%
ATLAS uses a right-handed coordinate system with its origin at the nominal interaction point (IP)
in the centre of the detector and the $z$-axis along the beam pipe.
The $x$-axis points from the IP to the centre of the LHC ring,
and the $y$-axis points upwards.
Cylindrical coordinates $(r,\phi)$ are used in the transverse plane,
$\phi$ being the azimuthal angle around the $z$-axis.
The pseudorapidity is defined in terms of the polar angle $\theta$ as $\eta = -\ln \tan(\theta/2)$.
Angular distance is measured in units of $\Delta R \equiv \sqrt{(\Delta\eta)^{2} + (\Delta\phi)^{2}}$.}

\label{sec:atlas1}
The ATLAS experiment~\cite{PERF-2007-01} at the LHC is a multipurpose particle detector
with a forward--backward symmetric cylindrical geometry and a near $4\pi$ coverage in
solid angle.\footnote{\AtlasCoordFootnote}
It consists of an inner tracking detector surrounded by a superconducting solenoid
providing a \SI{2}{\tesla} axial magnetic field, electromagnetic and hadronic calorimeters, and a muon spectrometer.
The inner tracking detector, covering the pseudorapidity range $|\eta| < 2.5$,
consists of silicon pixel and silicon microstrip tracking detectors inside a transition-radiation tracker that covers $|\eta| < 2.0$.
It includes, for the $\sqrt{s} = 13$~\TeV running period, a newly installed innermost pixel layer, the insertable $B$-layer~\cite{CERN-LHCC-2010-013}.
Lead/liquid-argon (LAr) sampling calorimeters provide electromagnetic (EM) energy measurements for $|\eta| < 2.5$
with high granularity and longitudinal segmentation.
A hadronic calorimeter consisting of steel and scintillator tiles covers the central pseudorapidity range ($|\eta| < 1.7$).
The endcap and forward regions are instrumented with LAr calorimeters for EM and hadronic energy measurements up to $|\eta| = 4.9$.
The muon spectrometer surrounds the calorimeters and is based on three large air-core toroid superconducting magnets with eight coils each.
It includes a system of precision tracking chambers ($|\eta| < 2.7$) and fast detectors for triggering ($|\eta| < 2.4$).
A two-level trigger system is used to select events~\cite{TRIG-2016-01}.
The first-level trigger is implemented in hardware and uses a subset of the detector information
to reduce the accepted rate to a design maximum of 100~kHz.
This is followed by a software-based trigger with a sustained average accepted event rate of about 1~kHz.

\section{Data and simulation}
\label{sec:Simulation}
\label{subsec:ds}
The data sample collected by the ATLAS Collaboration at $\sqrt{s}=13$~\TeV during 2015 and 2016 was used. 
After the application of beam and data quality requirements, the integrated luminosity 
is 36.1~\ifb.

Monte Carlo (MC) simulation samples were produced for signal and background processes using the full ATLAS detector simulation~\cite{SOFT-2010-01} based on \textsc{Geant4}~\cite{geant4} or, for selected smaller backgrounds and some of the signal samples, a fast simulation using a parameterisation of the calorimeter response and \textsc{Geant4} for the tracking system~\cite{ATL-PHYS-PUB-2010-013}.
To simulate the effects of additional $pp$ collisions in the same and nearby bunch crossings (pile-up), additional interactions were generated using \PYTHIA~8.186~\cite{Sjostrand:2006za,Sjostrand:2007gs} 
with a set of tuned parameters for the underlying event, referred to as the A2 tune~\cite{Hesketh:2041288}, and the MSTW2008LO set of parton distribution functions (PDF)~\cite{Martin:2009iq}, and overlaid on the simulated hard-scatter event. The simulated events were reweighted to match the distribution of the number of interactions per bunch crossing observed in the data and were reconstructed using the same procedure as for the data. 

The signal events containing \hpp pairs were simulated with the CalcHEP generator version 3.4~\autocite{Belyaev:2012qa}, which is at leading order in QCD,
using the Lagrangian described in Ref.~\cite{Arhrib:2011uy} and the PDF set CTEQ6L1~\cite{cteq6l1,cteq6}. The modelling of the parton showering and hadronisation of these events was performed using PYTHIA~8.186~\cite{Sjostrand:2006za,Sjostrand:2007gs} with the A14 tune~\cite{ATL-PHYS-PUB-2014-021}.
Event samples for the process \hppp were simulated for $m_{H^{\pm\pm}}$ in the range from $200$ to $700$ GeV with steps of $100$~\GeV. The production cross-section decreases rapidly with $m_{H^{\pm\pm}}$ and is $80.7$~fb for $m_{H^{\pm\pm}}=200~\GeV$, $5.0$~fb for $m_{H^{\pm\pm}}=400~\GeV$, and $0.35$~fb for $m_{H^{\pm\pm}}=700~\GeV$. 
Next-to-leading order (NLO) corrections~\cite{Muhlleitner:2003me} in QCD were applied, which increase these cross-sections by a factor 1.25. The fast detector simulation was used for the samples corresponding to $m_{H^{\pm\pm}} > 500$~\GeV.

\label{sec:mc}
\begin{table}[t]
\begin{center}
\caption{\label{tbl:evgen} Configurations used for event generation of background processes. If only one PDF is shown, the same is used for both the matrix element (ME) and parton shower generators; if two are shown, the first is used for the matrix element calculation and the second for the parton shower. $V$ refers to the production of an electroweak boson ($W$ or $Z/\gamma^*$). ``Tune'' refers to the underlying-event tune of the parton shower generator. ``\textsc{MG5\_aMC}'' refers to \textsc{MadGraph5\_aMC@NLO} 2.2.1; ``\textsc{Pythia} 6'' refers to version 6.427; ``\textsc{Pythia} 8'' refers to version 8.1; ``\Herwigpp'' refers to version 2.7. The samples 
have heavy flavour hadron decays modelled by \textsc{EvtGen}1.2.0~\cite{Lange:2001uf}, except for samples generated with \textsc{Sherpa}. 
}
{\footnotesize
\begin{tabular}{llllll}
\hline
Process & Event Generator & ME order & Parton Shower & PDF & Tune\\ 
\hline
\makecell[l]{$VV$, $qqVV$,\\$VVV$}& \textsc{Sherpa} 2.1.1 \cite{sherpa} & MEPS NLO & \textsc{Sherpa} 2.1.1  & CT10  \cite{ct10} & \textsc{Sherpa} 2.1.1 default \\
\makecell[l]{\ttH }& \textsc{MG5\_aMC} \cite{Alwall:2014hca} & NLO & \textsc{Pythia} 8 \cite{Sjostrand:2007gs}\ & NNPDF 3.0 NLO \cite{Ball:2014uwa} & A14 \cite{ATL-PHYS-PUB-2014-021} \\
\makecell[l]{$VH$ }&\PYTHIAV  8 &  LO &  \textsc{Pythia} 8 &  NNPDF 2.3 LO &  A14 \\  
\makecell[l]{$tHqb$ }& \textsc{MG5\_aMC} & LO &\textsc{Pythia} 8 & CT10 & A14\\
\makecell[l]{$tHW$ }& \textsc{MG5\_aMC} & NLO & \Herwigpp \cite{Bahr:2008pv} & CT10 & UE-EE-5   \cite{Seymour:2013qka} \\
\makecell[l]{$\ttbar W$, $\ttbar (Z/\gamma^*)$}& \textsc{MG5\_aMC} & NLO & \textsc{Pythia} 8 & NNPDF 3.0 NLO & A14  \\
\makecell[l]{$t (Z/\gamma^*)$ }&  \textsc{MG5\_aMC} & LO & \textsc{Pythia} 6 \cite{Sjostrand:2006za} & CTEQ6L1~\cite{cteq6l1,cteq6}  & Perugia2012 \cite{perugia} \\
\makecell[l]{$t W (Z/\gamma^*)$ }& \textsc{MG5\_aMC} & NLO &  \textsc{Pythia} 8 & NNPDF 2.3 LO  & A14 \\
\makecell[l]{$\ttbar t,  t\bar t t\bar t$ }& \textsc{MG5\_aMC} & LO & \textsc{Pythia} 8 & NNPDF 2.3 LO & A14\\
\makecell[l]{$t\bar t W^+ W^-$ }& \textsc{MG5\_aMC} & LO & \textsc{Pythia} 8 & NNPDF 2.3 LO & A14\\
\makecell[l]{$V\gamma $ }& \textsc{Sherpa} 2.2 & MEPS NLO& \textsc{Sherpa} 2.2  & NNPDF 3.0 NLO & \textsc{Sherpa} 2.2 default \\
\makecell[l]{$s$-, $t$-channel,\\ $Wt$ single top} & \POWHEGBOXV v2 \cite{powhegstp,powhegstp2} & NLO & \textsc{Pythia} 6 & CT10/CTEQ6L1 & Perugia2012 \\
\hline
\end{tabular}
}

\end{center}
\end{table}

The SM background processes were simulated using the MC event generator programs and configurations shown in Table~\ref{tbl:evgen}.  
The production of $VV$, $VVqq$, and $VVV$ (where $V$ denotes a vector boson $W$ or $Z$ and $qq$ labels the vector-boson fusion production mechanism) was simulated with a NLO 
QCD matrix element computed by \textsc{Sherpa} and matched to the \textsc{Sherpa} parton shower. 
The main background contribution in the \ll and \lll channels is from  $WZ$ production, for which the total cross-section prediction is $48.2\pm1.1$~pb~\cite{Grazzini:2016swo}. 
The main contribution to the \llll topology is from $ZZ$ production with a total cross-section of $16.9\pm0.6$~pb~\cite{Caola:2015psa, Grazzini:2015hta}, which is suppressed by requiring significant missing transverse momentum in these events.
The MC samples used to simulate \ttH, \ttV, $VV$ and \ttbar are described in more detail in Refs.~\cite{ATL-PHYS-PUB-2016-005,ATL-PHYS-PUB-2017-005,ATL-PHYS-PUB-2016-001}.

The simulated SM contributions in each of the channels considered are separated into prompt-lepton and fake-lepton contributions, depending on the source of the reconstructed leptons at generator level. The processes that contain only reconstructed charged leptons originating from prompt leptonic 
decays of $W$ and $Z$ bosons are classified as a prompt-lepton contribution, while processes with at least 
one of the reconstructed leptons being a misidentified hadron or photon, or
a lepton from hadron decays constitute the fake-lepton contribution.
The simulated events are not used to evaluate the background originating from charge-misidentified leptons for the \ll channel and fake leptons for the \ll and \lll channels. These are estimated in general using data-driven methods because they are not well modelled by simulations. This is the case in
particular for  $Z \to \ell^+\ell^-$, $W \to \ell \nu$ and \ttbar processes.
The background process $V\gamma$ can contribute if electrons originating from the photon conversion are selected. This contribution is found to be small and adequately modelled, so it is estimated using the MC simulation. For the \llll channel, the background from fake leptons is small and the data-driven methods are not applicable due to the low number of events available, so the MC simulation is used to estimate both the prompt-lepton and the fake-lepton contributions.

\section{Event selection and background estimates}
\label{sec:EventSelection}
\subsection{Event reconstruction}
Interaction vertices originating from $pp$ collisions are reconstructed using at least two tracks with transverse momentum $\pt > 0.4$~\GeV, and required to be consistent with the beam-spot envelope. 
The primary vertex is identified as the vertex with the largest sum of squares of the transverse momenta from associated tracks~\cite{ATL-PHYS-PUB-2015-026}.

Electrons are reconstructed as tracks in the inner detector matched to clusters in the electromagnetic calorimeter, within the region of pseudorapidity $|\eta| <  2.47$~\cite{ATLAS-CONF-2016-024}. The candidates in the transition region between the barrel and the endcap calorimeters ($1.37  < |\eta| <  1.52$) are removed. Only those electron candidates with transverse momentum greater than 10~\GeV
are considered. The electron identification is based on a multivariate likelihood-based discriminant that uses the shower shapes in the electromagnetic calorimeter and the associated 
track properties measured in the inner detector. 
In particular, the \textit{loose} and \textit{tight} identification working points, described in Ref.~\cite{ATLAS-CONF-2016-024}, are used, providing electron identification efficiencies of approximately $95$\% and $78\%$--$90\%$ (depending on \pt and $\eta$),
respectively. In order to reduce contributions from converted photons and hadron decays, the longitudinal impact parameter of the electron
track relative to the selected event primary vertex, multiplied by the sine of the polar angle, $|z_{0}\sin\theta|$, is required to be 
less than 0.5 mm. The transverse impact parameter divided by its uncertainty, $|d_{0}|/\sigma(d_{0})$, is 
required to be less than five. The identification algorithm is complemented by an isolation requirement, based on the energy 
in a cone around the electron candidate calculated using either charged tracks or calorimetric deposits. The calorimeter- and track-based 
isolation criteria are applied jointly to suppress fake electrons. 

 Muon candidates are reconstructed by combining tracks formed in the inner detector and in  the muon spectrometer, within 
the region of pseudorapidity $|\eta| < $ 2.5~\cite{PERF-2015-10}. Only those muon candidates with transverse momentum greater than 
10~\GeV are considered. A muon candidate is required to satisfy \textit{loose} or \textit{tight} identification criteria which are defined in Ref.~\cite{PERF-2015-10}, and which have efficiencies of approximately 98\% and 92\%, respectively. Similarly to electrons, isolation criteria complement the identification requirements. The impact parameters must satisfy $|z_{0}\sin\theta| < 0.5$~mm and $|d_{0}|/\sigma(d_{0}) < 3$ when selecting muons.

Combining the selection criteria mentioned above, two types of lepton requirements are used for both the electrons and muons: type $T$ (for \textit{tight}) and $L$ (for \textit{loose}). The type $T$ leptons are a subset of the type $L$.

Jets are reconstructed from topological clusters~\cite{PERF-2014-07} of energy deposits in the calorimeters using the anti-$k_t$ algorithm~\cite{Cacciari:2008gp,Cacciari:2011ma} with a radius parameter of $R = 0.4$. Only jets with $\pt > 25$~\GeV and $|\eta| < 2.5$ are considered. In order to suppress jets arising from pile-up collisions, jets with $\pt < 60$~\GeV and $|\eta| <2.4$ must have a sizeable fraction of their tracks matched to the selected primary vertex~\cite{PERF-2014-03}. 
Jets containing $b$-hadrons are identified ($b$-tagged) via a multi-variate discriminant combining information from the impact parameters of displaced tracks with topological properties of
secondary and tertiary decay vertices reconstructed within the jet~\cite{PERF-2012-04}.
 The $b$-tagging algorithm used for this search has an average efficiency of 70\% to identify $b$-jets with $\pt > 20$~\GeV and $|\eta| < 2.5$ in simulated $t\bar t$ events. 

To avoid object double counting, an overlap removal procedure is applied to resolve ambiguities among electrons, muons, and jets in the final state. 
Any electron candidate sharing an inner detector track with a muon candidate is removed. 
Jets within $\Delta R = 0.2$ of an electron, as well as  jets with less than three tracks within  $\Delta R = 0.2$ of a muon candidate are discarded. Any remaining electron candidate within  $\Delta R = 0.4$ of a jet
is discarded. Any remaining muon candidate within $\Delta R = 0.04 + 10/\pt^{\mu} (\textrm{GeV})$ of a jet is discarded. 
 
The missing transverse momentum, with magnitude $E_\text{T}^\text{miss}$, is defined as the negative vector sum of the
transverse momenta of all identified leptons and jets and the remaining unclustered energy
of the event, which is estimated from tracks associated with the primary vertex but not 
assigned to any physics object~\cite{PERF-2016-07}. 

\subsection{Event preselection}

Candidate events are selected using triggers that require at least one electron or one muon to pass various thresholds of $\pt$~\cite{TRIG-2016-01}. The higher thresholds are applied with looser lepton identification and/or isolation requirements in order to ensure efficiencies close to 100\% for leptons with transverse momentum above 30~\GeV.

The signal topologies studied in this search involve the presence of at least two leptons of the same charge and are classified
as explained above in three mutually exclusive categories: \ll, \lll and \llll channels. 
The \ll channel targets signal events where the two same-sign $W$ bosons from one of the doubly charged Higgs boson decays leptonically, while the two $W$ bosons from the other doubly charged Higgs boson decay hadronically. 
In the \lll channel, one $W$ boson decays hadronically and in the \llll channel, all $W$ bosons decay leptonically. All channels present significant \MET corresponding to the neutrinos from leptonic $W$ boson decays.  In the \ll and \llll channels, jets from $W$ boson decays originate from the first- and second-generation quarks, and therefore lead to events without $b$-jets. 
The event selection is divided into two steps: the preselection and the signal region selection. 

The preselection requirements are summarised in Table~\ref{Tab:preselection}. The electrons (muons) are selected in the pseudorapidity range $|\eta| < 2.47\ (2.5)$ with a transverse momentum of at least 10~\GeV, satisfying the type $L$ requirement. 
Events are selected only if the absolute value of the sum of charges of the leptons is two, one and zero for the ${2\ell^{ss}}$,  \lll and \llll channels, respectively.  
At least one of the leptons is required to have \pt $> 30$~\GeV to ensure a high trigger efficiency. 
To reduce the fake-lepton contamination in the \ll channel, the second highest \pt (subleading) lepton  is required to have \pt  $> 20$~\GeV and both leptons are required to be of type $T$. 
Similarly in the \lll channel, each lepton in the pair of leptons of the same sign, which is expected to suffer more from fake-lepton contamination, is required to have \pt $> 20$~\GeV and to both be of type $T$. 
In the \ll and \llll channels,  the leptons are labelled by descending \pt, and are denoted by $\ell_{1,2,...}$. 
The ranking follows a different logic for the \lll channel: the lepton that has a charge opposite to the total lepton charge is denoted as  $\ell_0$, while the same-sign leptons are denoted by $\ell_1$ and $\ell_2$, ranked by increasing distance to $\ell_0$ in the $\eta$--$\phi$ plane. 

Further preselection requirements are based on \MET,  the jet multiplicity \nj and the number of jets tagged as $b$-jets $N_{b\textrm{-jet}}$. 
Moreover, in order to reduce the background from $Z$ bosons and neutral mesons decaying into same-flavour opposite-sign leptons (SFOS), the invariant mass of such lepton pairs is required to be greater than 12 (15)~\GeV for the \lll (\llll) channel and to have an invariant mass that is not compatible with the $Z$ boson. For the \ll channel, the $Z$ boson invariant mass veto is also applied  to $e^\pm e^\pm$ events, 
in order to reduce the contributions originating from electron charge misidentification. 

 After this preselection, $562$ data events are selected in the \ll channel, $392$ events in the \lll channel, and $44$ events in the \llll channel. 

\begin{table}[htbp]
\footnotesize
\renewcommand{\arraystretch}{1.5}
 \begin{center}
 \caption{\label{Tab:preselection} The preselection criteria for the three analysis channels. The leptons are ordered by decreasing $\pt$ ($\ell_1,\ell_2,...$) in the \ll and \llll channels, while for the \lll channel $\ell_1,\ell_2$ denote the same-sign leptons and $\ell_0$ the lepton with a charge opposite to the total lepton charge. $Q_\ell$ denotes the charge of each lepton.} 
   \begin{tabular}{l|c|c|c}
Selection criteria & \ll & \lll & \llll \\
\hline
Trigger & \multicolumn{3}{c}{At least one lepton with $\pt^{\ell}>30$~\GeV } \\
    & \multicolumn{3}{c}{that fulfils the requirements of single-lepton triggers } \\
\hline
$N_\ell$($L$-type, $\pt>10$~\GeV, $|\eta_{\ell}|<2.47$) &2 & 3& 4\\ 
$N_\ell$($T$-type, $\pt>10$~\GeV, $|\eta_{\ell}|<2.47$) & 2 & 2 ($\ell_{1,2}$) & $-$ \\ 
$|\sum Q_\ell|$ & 2 & 1& 0 \\
Lepton $\pt$ threshold & $\pt^{\ell_1,\ell_2}>30, 20$~\GeV &    $\pt^{\ell_0,\ell_1,\ell_2}>10, 20, 20$~\GeV  &  $\pt^{\ell_1,\ell_2,\ell_3,\ell_4}>10$~\GeV \\  
\hline

\hline
 \MET &  $>70$~\GeV & $>30$~\GeV &  $>30$~\GeV \\
\hline
 \nj  &  $\ge 3$ & $\ge 2$ &  -- \\
 \hline
$b$-jet veto  & \multicolumn{3}{c}{$N_{b \text{-jet}}=0$ } \\
\hline
 Low SFOS $m_{\ell\ell}$ veto & --  & $m_{\ell^\pm\ell^\mp}>15~\GeV$ & $m_{\ell^\pm\ell^\mp}>12~\GeV$  \\
\hline
 $Z$ boson decays veto &  $|m_{e^\pm e^\pm}-m_Z|>10~\GeV$  & \multicolumn{2}{c}{$|m_{\ell^\pm\ell^\mp}-m_Z|>10~\GeV$ } \\
\hline
   \end{tabular}
  
 \end{center}
\end{table}

\subsection{Background estimate}

The background processes containing only prompt selected leptons are estimated with MC simulations normalised to the most precise cross-section calculation (see Section~\ref{sec:mc}). Further contributions originate from non-prompt and  mismeasured leptons. 
The procedures used to estimate those contributions are described in the following.

\subsubsection{Charge misidentification} 
\label{sec:chargeflip}
In the \ll channel, a background contribution is expected from events with opposite-sign lepton pairs when the charge of one of the
leptons is misidentified, while the background contribution from charge misidentification is negligible for \lll and \llll  channels. 
In the transverse momentum domain relevant for this analysis, charge misidentification is only significant for electrons and is due mainly to bremsstrahlung interactions with the inner detector material. The radiated photon produces an $e^+e^-$ pair near the original electron trajectory leading to a charge identification confusion.

The misidentification rate is measured using a large data sample of
dilepton events originating mainly from $Z\rightarrow e^+e^-$ decays selected by two type $T$ electrons with an invariant mass between 80 and 100~\GeV. 
The sample contains mostly opposite-sign dileptons, with a small fraction of same-sign dileptons. The fraction of same-sign dilepton events is used to extract the charge-misidentification rate 
as a function of  electron \pt and $\eta$. 
This rate is found to range between $0.02$\% and $10$\%, where large values
are obtained at large rapidities where the amount of material is higher. The statistical error of this estimate is taken as systematic uncertainty of the charge misidentification rate.
The background from fake leptons in both the opposite-sign and same-sign samples is estimated using sidebands around the $Z$ boson mass peak. 
Its impact on the charge misidentification rate is about 2\% and is included in the systematic uncertainty.

The background from charge misidentification in a given region is estimated using a data control sample selected with the same criteria as the nominal sample but with opposite-sign dilepton pairs, where at least one lepton is an electron, weighted by the
probability that the charge of the electron(s) is misidentified. 

\subsubsection{Fake-lepton contributions}
\label{sec:fakeLeptons}
The composition of the fake-lepton background varies considerably among the analysis channels. Therefore, 
the methods to estimate the fake-lepton contributions are different for the \ll, \lll and \llll channels. 
The contribution from fake leptons for the \ll and \lll channels are estimated using the fake-factor method, while the simulation prediction corrected with data-driven scale factors is used for the \llll channels.    
Those methods involve various fake-enriched control samples that are summarised in Table~\ref{Tab:ControlFakes} and described below.

\begin{table}[!htbp]
 \begin{center}
 \caption{\label{Tab:ControlFakes} The selection criteria defining the fake-enriched control regions used to determine the fake factors for the \ll and \lll channel 
 and the MC scale factors for the \llll channel. The symbol "$-$" means no
 requirement. The transverse mass $m_\text{T}$, used for the  $4\ell$-Z region to reduce the $WZ$ contributions, is calculated as the invariant mass of the vector sum of transverse momentum of the fake-lepton candidate and the missing transverse momentum.
} 
 \tabcolsep=0.11cm
 \footnotesize
   \begin{tabular}{ l| c |c |c |c }
\hline
Sample     & \ll & \lll        & \llll-Z            & \llll-T\\
\hline
$N_{\ell}$ (type $L$)      & 2            & 3               & 3                     & 3         \\
$|\sum Q_\ell|$ & 2 & 1& 1 &1 \\
$\pt^{\ell}$ & $>30,20$~\GeV & $>10,20,20$~\GeV & $>10,10,10$~\GeV       & $>10,10,10$~\GeV\\
\nj     &  $\ge 3$         & $1$           & $1 \; \mathrm{or} \;2 $          & $1 \; \mathrm{or} \; 2 $ \\
$N_{b \text{-jet}}$  & $0$          & $-$           & $-$                  & $-$\\
$\pt^{jet}$  & $>25$~\GeV     & $>25$~\GeV      & $>25$~\GeV             & $>30 (25)$~\GeV\\
$Z$-window     & $|m_{ee}^\text{ss}- m_{Z}| > 10$~\GeV &$|m_{\ell\ell}^\text{os}- m_{Z}| > 10$~\GeV &$|m_{\ell\ell}^\text{os}- m_{Z}| < 10$~\GeV & No same-flavour\\
               &               &                &                       & opposite-sign lepton pair\\
$m_{\ell\ell}^\text{os}$& $-$      & $>15$~\GeV      & $-$                  & $-$\\
$\MET$         & $<70$~\GeV     & $-$           & $<50$~\GeV             & $-$\\
$m_\text{T}$       & $-$          & $-$           & $<50$~\GeV             & $-$ \\
\hline

\end{tabular}
\end{center}
  
\end{table}

\paragraph{Fake-lepton contribution estimate for the \ll channel}

The fake-factor method assumes that the fake-lepton contribution in a nominal region, which can be the preselection or the signal region, can be computed using an extrapolation factor that is referred to as a fake factor, and is denoted as $\theta$ in the following. The fake factor is multiplied by the number of events containing fake leptons in a region with the same selection criteria as the nominal region, except that at least one of the leptons is required to satisfy the type $L$ but not the type $T$ identification criteria. That lepton is denoted by $\cancel{e}$, $\cancel{\mu}$ or collectively $\cancel{\ell}$ in the following. 

The fake factors are calculated in fake-enriched control regions with kinematic selections designed to enhance their content in fake leptons. In the case of the  \ll channel (\ll column in Table~\ref{Tab:ControlFakes}), this is achieved by requiring low $\MET$.
The fake factor is defined as the number of fake-lepton events in the fake-enriched region where all selected leptons pass the type $T$ identification, divided by the number of fake-lepton events in the same region but where one of the selected leptons is of type $\cancel{\ell}$.

The muon fake factor is thus computed in the fake-enriched region, where a pair of same-sign muons was selected, as follows:

\begin{equation}
 \theta^{2\ell^\text{ss}}_{\mu} =\frac{(N^\mathrm{Data}-N^\mathrm{Prompt})^C_\mathrm{\mu\mu}} {(N^\mathrm{Data}  - N^\mathrm{Prompt})^C_\mathrm{\mu\cancel{\mu}} }, \label{eq:ffcalc_2L_mu}
\end{equation}

where $N^\mathrm{Data, C}_\mathrm{\mu\mu}$  and $N^\mathrm{Data, C}_\mathrm{\mu\cancel{\mu}}$ are the number of events where both muons are of type $T$, and where one is of type $T$ and the other of type $\cancel{\mu}$, respectively. 
The prompt-lepton contributions $N^\mathrm{Prompt}$, which are estimated using MC simulation, are subtracted from data event yields to obtain a pure estimate of the fake-lepton contributions in the $\mu\mu$ and $\mu\cancel{\mu}$  regions. The superscript $C$ indicates the fake-enriched control region. 

The electron fake factor is computed using the fake-enriched region where a same-sign $e\mu$ pair was selected:

\begin{align}
 \theta^{2\ell^\text{ss}}_{e} &=\frac{(N^\mathrm{Data}-N^\mathrm{Prompt} - N^\mathrm{\cancel{Q}} -  N^\mathrm{fake\text{-}\mu}
 )^C_{e\mu}} {(N^\mathrm{Data}  - N^\mathrm{\cancel{Q}} - N^\mathrm{Prompt})^C_{\cancel{e}\mu} }. \label{eq:ffcalc_2L_e}
\end{align}

In addition to the prompt-lepton contribution, the electron charge-misidentification contribution, denoted by $N^\mathrm{\cancel{Q}}$, needs to be subtracted. It is computed using the method described in Section~\ref{sec:chargeflip}. Furthermore, the fake-muon contribution in the $e\mu$ sample is subtracted from the numerator. It is computed as:

\begin{align}
 N^{\mathrm{fake\text{-}\mu},C}_{e\mu} &= \theta^{2\ell^\text{ss}}_{\mathrm \mu}  (N^\mathrm{Data}  - N^\mathrm{Prompt})^C_{e \cancel{\mu}}. \nonumber  \end{align}

The fake-muon contribution is not considered in the denominator of the electron fake factor, in Eq.~(\ref{eq:ffcalc_2L_e}), because it is negligible. 

The muon fake factor is measured to be  $0.14\pm0.03$, while the electron fake factor
is $0.48\pm0.07$, where the uncertainties are statistical only. A systematic uncertainty of 35\% (56\%) in the electron (muon) fake factor is estimated from complementary control samples with low jet multiplicity or by applying a different selection to vary the fraction of jets containing heavy-flavour hadrons. 
The uncertainty in the muon fake factor is larger than in the electron fake factor due lower number of data events available for those checks.
The fake-lepton contributions in the nominal region (signal or preselection, denoted collectively by the superscript $R$) are obtained by multiplying the fake factors by the number of events in a region with the same selection as the nominal region, but where at least one lepton is of type $\cancel{\ell}$: 

\begin{align}
N^{\mathrm{fake},R}_{e\mu} &= \theta^{2\ell^\text{ss}}_{e}     \times (N^\mathrm{Data}-N^\mathrm{Prompt} - N^\mathrm{\cancel{Q}} )^R_{\mu \cancel{e}} 
                                               +\theta^{2\ell^\text{ss}}_{\mu} \times  (N^\mathrm{Data}-N^\mathrm{Prompt} - N^\mathrm{\cancel{Q}} )^R_{e\cancel{\mu}}  \nonumber \; \; , \\
N^{\mathrm{fake},R}_{ee,\; \mu\mu} &= \theta^{2\ell^\text{ss}}_{e,\mu} \times (N^\mathrm{Data}-N^\mathrm{Prompt} - N^\mathrm{\cancel{Q}} )^R_{e\cancel{e},\; \mu\cancel{\mu}}  \label{eq:ff_ll} \; . 
\end{align}

where the prompt-lepton and the charge misidentifications contributions are subtracted as explained above. 

\paragraph{Fake-lepton contribution estimate for the \lll channel}
A method similar to that employed for the \ll channel is applied for the \lll channel. Here the opposite-sign lepton $\ell_0$ is assumed to be prompt, an assumption that was found to be valid in MC simulation. 
The fake-enriched region used to calculate the fake factors for the \lll channel, which is described in Table~\ref{Tab:ControlFakes}, follows the \lll preselection conditions except that the jet multiplicity is required to be exactly one. 
The fake factors for electrons and muons are both calculated by applying a formula analogous to Eq.~(\ref{eq:ffcalc_2L_mu}) to the $\ell_0 ee/ \ell_0 e\cancel{e}$ and $\ell_0 \mu\mu/ \ell_0 \mu\cancel{\mu}$ regions, respectively.
The muon fake factor is found to be $0.17\pm0.06$ and the electron fake factor is found to be $0.39\pm0.07$, where the errors are statistical only. The values are compatible with those obtained for the \ll channel. Additional control samples, defined such that the content is enriched in either \zj or \ttbar events, are used to test the method and to estimate systematic uncertainties of 55\% and 81\% for the electron and muon fake factors, respectively. The fake-lepton contributions to the nominal regions are then calculated using  relations analogous to Eq.~(\ref{eq:ff_ll}).

\paragraph{Fake-lepton contribution estimate for the \llll channel}
There are too few data events to apply the fake-factor method in the \llll channel. Instead, the fake-lepton contribution is estimated from the yields predicted by the MC simulation 
but corrected using process-dependent scale factors that are extracted in two fake-enriched
control regions. 
The fake-lepton contribution in this channel comes mainly from \ttV processes, where the fake lepton originates from a $b$-jet.
A small component from light quarks is also present. 
Two data samples designed to contain fake leptons originating from \zj and \ttbar 
events are used to study the capability of the simulation to describe fake leptons originating from light- and heavy-flavour jets, respectively. The two control samples are labelled Z and T and are defined in Table~\ref{Tab:ControlFakes}. The samples are required to have three identified leptons. For the Z region, the fake-lepton candidate is assumed not to be part of the lepton pair forming the $Z$ boson candidate. For the T region, the fake lepton is assumed to be the lepton with the lower \pt in the same-sign lepton pair. The scale factors are derived independently for fake electrons and fake muons. 
Four scale factors  $\lambda_\mathrm{X}^{\ell}$ (with $\ell=e,\mu$ and $\mathrm{X=Z,T}$)
are obtained by solving the system of equations

\begin{equation}
 N^{\ell}_\mathrm{Data | X}-N^{\ell}_\mathrm{Prompt | X}  = \lambda_\text{T}^{\ell}N^{\ell}_{\ttbar | \mathrm{X}} + \lambda_\text{Z}^{\ell}N^\ell_{\zj | \mathrm{X}}, \nonumber
\end{equation}

where the event yields $N^{\ell}$ are labelled by the nature of the contribution, data (Data) or simulation (Prompt, \ttbar and \zj), and the equations are derived in each of the respective control region X (Z or T).
The obtained scale factors are 
$\lambda_\text{T}^{e} = 1.12 \pm 0.05 $, 
$\lambda_\text{Z}^{e} = 1.02 \pm 0.07 $,
$\lambda_\text{T}^{\mu} =1.11 \pm 0.05 $ and 
$\lambda_\text{Z}^{\mu} = 0.94 \pm 0.07$, where the errors are statistical only. Alternative trilepton control samples, where the jet multiplicity and the lepton \pt threshold are varied, are used to estimate a systematic uncertainty of $50$\% in these scale factors. 
The scale factors are used as weights to the simulated events that contain a fake lepton according to the fake-lepton flavour and the presence of heavy-flavour jets in the event.  

\section{Signal region optimisation}
\label{sec:SignalRegion}

The hypothetical signal produces four $W^\pm$ bosons in each event. Since at least two leptonic $W$ boson decays are 
needed to lead to the multi-lepton topologies considered in this analysis, all signal events are expected 
to feature significant \MET, while jets are expected from hadronic $W$ boson decays for \ll and \lll channels.
 Moreover, when the mass of the doubly charged Higgs boson is in the range of $200$--$300$~\GeV, each $\hpp$ is produced with a significant momentum and the two subsequent $W$ bosons are emitted close to each other in the laboratory frame. Consequently, the two same-sign leptons from the decays of the two $W$ bosons tend to be close in the $\eta$--$\phi$ plane. The decay products of the
 other doubly charged Higgs boson are generally well-separated from the two same-sign leptons. 

The analysis channels face different background contributions from the SM. The \ll
category is populated with events 
containing one prompt lepton from a $W$ boson, or to a lesser extent from a $Z$ boson, and one fake lepton from the hadronic final state produced.
The \ll events with two same-sign 
electrons can also originate from Drell--Yan and $t\bar{t}$ production, 
where the charge of one of the electrons is misidentified, as explained above. 
In the \ll and \lll channels, most of the expected prompt-lepton contribution is due to the production of $WZ$ associated with jets, 
with both bosons subsequently decaying into leptons. 
This process also produces other features of the signal, such as significant $E_\text{T}^\text{miss}$ and the absence of $b$-jets for most of the production cross-section. 
For the $WZ$ events, the mass of the same-flavour opposite-sign lepton pair is close to the $Z$ boson mass, while no such resonant distribution is expected for the signal. 
In the \llll channel, the dominant background originates from $t\bar{t}V$ and $ZZ$ production. Processes containing top quarks 
($t\bar{t}$, $t\bar{t}V$) can lead to events with multiple leptons in the final state. A noticeable feature 
of those processes is the presence of $b$-jets.

Given these properties of the signal and of the expected background, the following discriminating variables, in addition to \MET, are considered: \label{sec,var}
\begin{itemize}
\item $m_{x\ell}$, the invariant mass of the system composed of all selected leptons in the event, where $x$ can be 2, 3 or 4.
\item $\Delta R_{\ell^\pm\ell^\pm}$, the distance in $\eta$--$\phi$ between two same-sign leptons. 
This variable is used for the \ll and \lll channels. In the \llll channel, two such variables can be calculated per event, $\Delta R^\mathrm{min}_{\ell^\pm\ell^\pm}$ and $\Delta R^\mathrm{max}_{\ell^\pm\ell^\pm}$, denoting the minimum and maximum values, respectively. 
\item $m_\mathrm{jets}$, the invariant mass of the system composed of all jets in the event. When there are more than four jets in the event, only the leading four jets are used. This variable is used only for the \ll channel.
\item $ \pt^{\text{leading jet}}$, the transverse momentum of the highest-$\pt$ jet.  
\item $\Delta \phi({\ell\ell,\MET})$, the difference in azimuth between the dilepton system and $\MET$. This variable is used in the \ll channel. 
\item $\Delta R_{\ell-\mathrm{jet}}$, the minimal distance in $\eta$--$\phi$ between any lepton and its closest jet. This variable is used in the \lll channel.
\item $S$, is a variable used for the \ll channel to describe the event topology in the transverse plane, and defined using the spread of the $\phi$ angles of the leptons, $\MET$, and jets as follows: 
\begin{equation}
S = \frac{{\cal{R} }(\phi_{\ell_1},\phi_{\ell_2},\phi_{\MET})\cdot{\cal{R} }(\phi_{j1},\phi_{j2},\cdots  )}{{\cal{R} }(\phi_{\ell_1,},\phi_{\ell_2},\phi_{\MET},\phi_{j1},\phi_{j2},\cdots) }, \nonumber
\end{equation}
where the $\cal{R} $ is the root mean square that quantifies the spread, ${\cal{R}} (\phi_1, \cdots, \phi_n ) = \sqrt{\frac{1}{n} \sum\limits_{i=1}^{n} (\phi_i - \overline{\phi} )^2}$. The azimuthal angles $\phi$  are bounded in $(-\pi, \pi]$, and the bound is considered in the calculation. The $S$ variable is expected to be on average smaller for the signal than for the background for low \hpp mass values.
\end{itemize}

The distributions of the selected variables for the \ll, \lll and \llll channels are shown at preselection level in Figures~\ref{fig,2L_PRES},~\ref{fig,3L_PRES} and~\ref{fig,4L_PRES}, respectively. 
The data are compared with the sum of the prompt lepton, fake lepton and charge-misidentified lepton background predictions. The prompt-lepton backgrounds are estimated with simulations while the background from fake leptons and charge-flipped leptons are measured with the methods described in the previous section. Good agreement is observed in both 
normalisation and shape, demonstrating that the background contributions are well modelled. The expected signal distributions for various \hpp masses are also shown to illustrate the discriminating power of the selected variables. 

\begin{figure}[htp]
  \centering
  \includegraphics[scale=0.746]{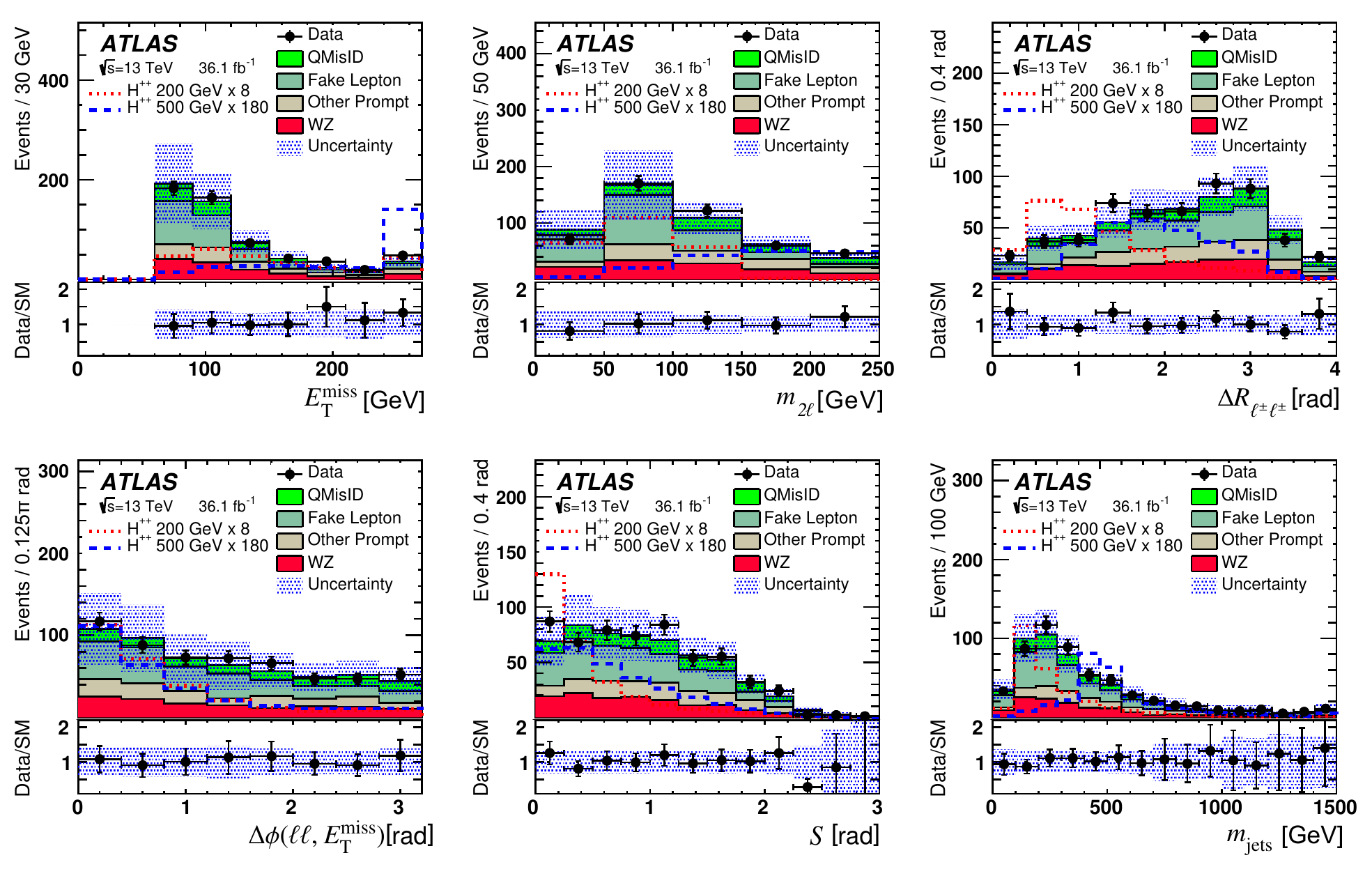}
  \caption{\label{fig,2L_PRES}Distribution of variables used for the signal region optimisation of the \ll final state. The events are selected with the preselection requirements listed in Table~\ref{Tab:preselection}.  The data (dots) are compared with the predictions (histograms) that include the contributions from the dominant prompt-lepton background ($WZ$), other prompt-lepton backgrounds, processes where a fake lepton is reconstructed, and electrons with misidentified charge (QMisID). The expected signal distributions corresponding to two \hpp masses are also shown, scaled up for visibility. The last bin includes overflows. In each figure the bottom panel shows the ratio of data to the prediction, where the band around unity represents the total uncertainty of the SM prediction.}
\end{figure}

\begin{figure}[htp]
  \centering
  \includegraphics[scale=0.746]{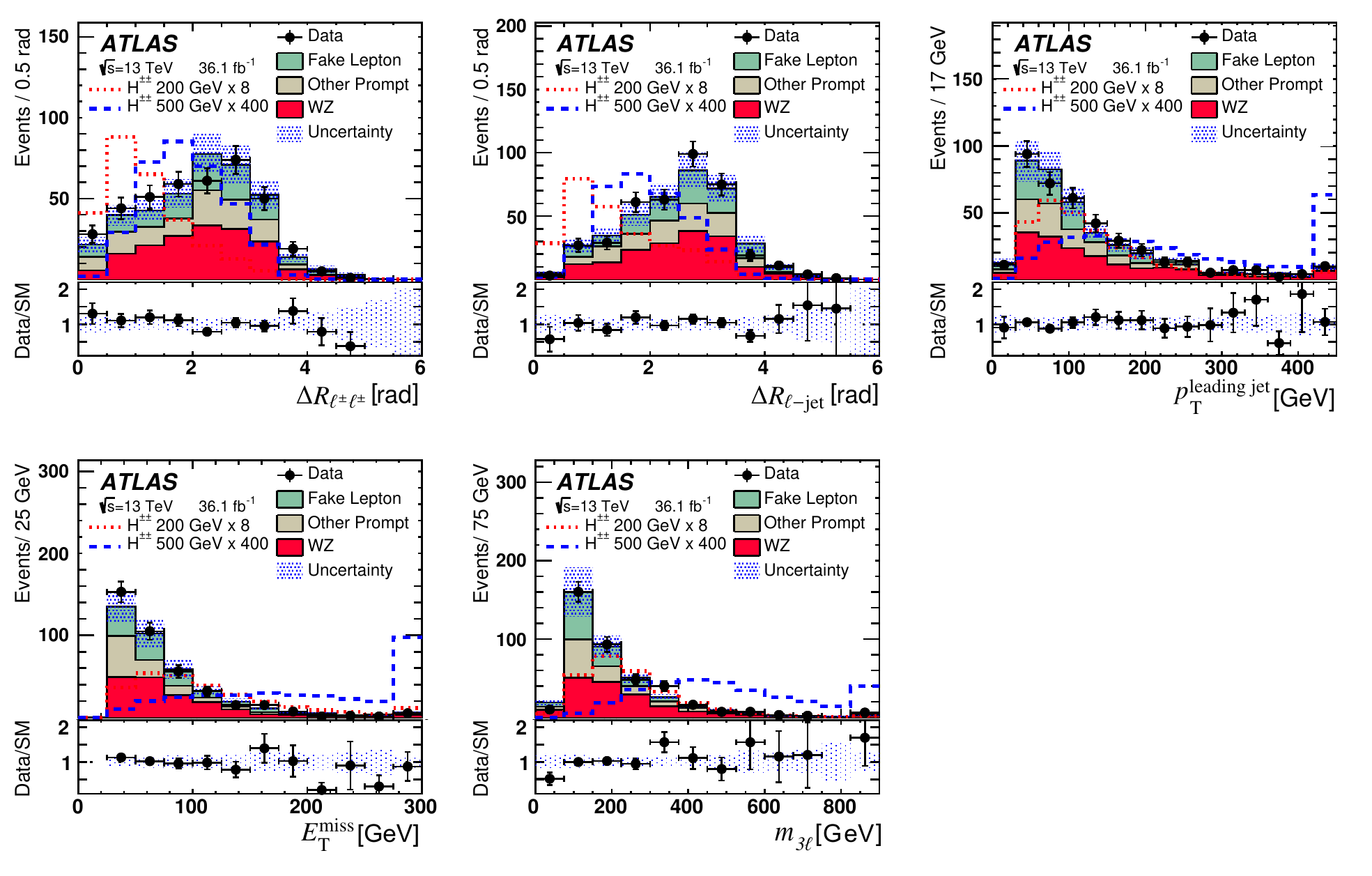}
  \caption{\label{fig,3L_PRES} Distribution of variables used for the signal region optimisation of the \lll channel (a detailed description can be found in the caption of Figure~\ref{fig,2L_PRES}).}
\end{figure}

\begin{figure}[htp]
        \centering
        \includegraphics[scale=0.746]{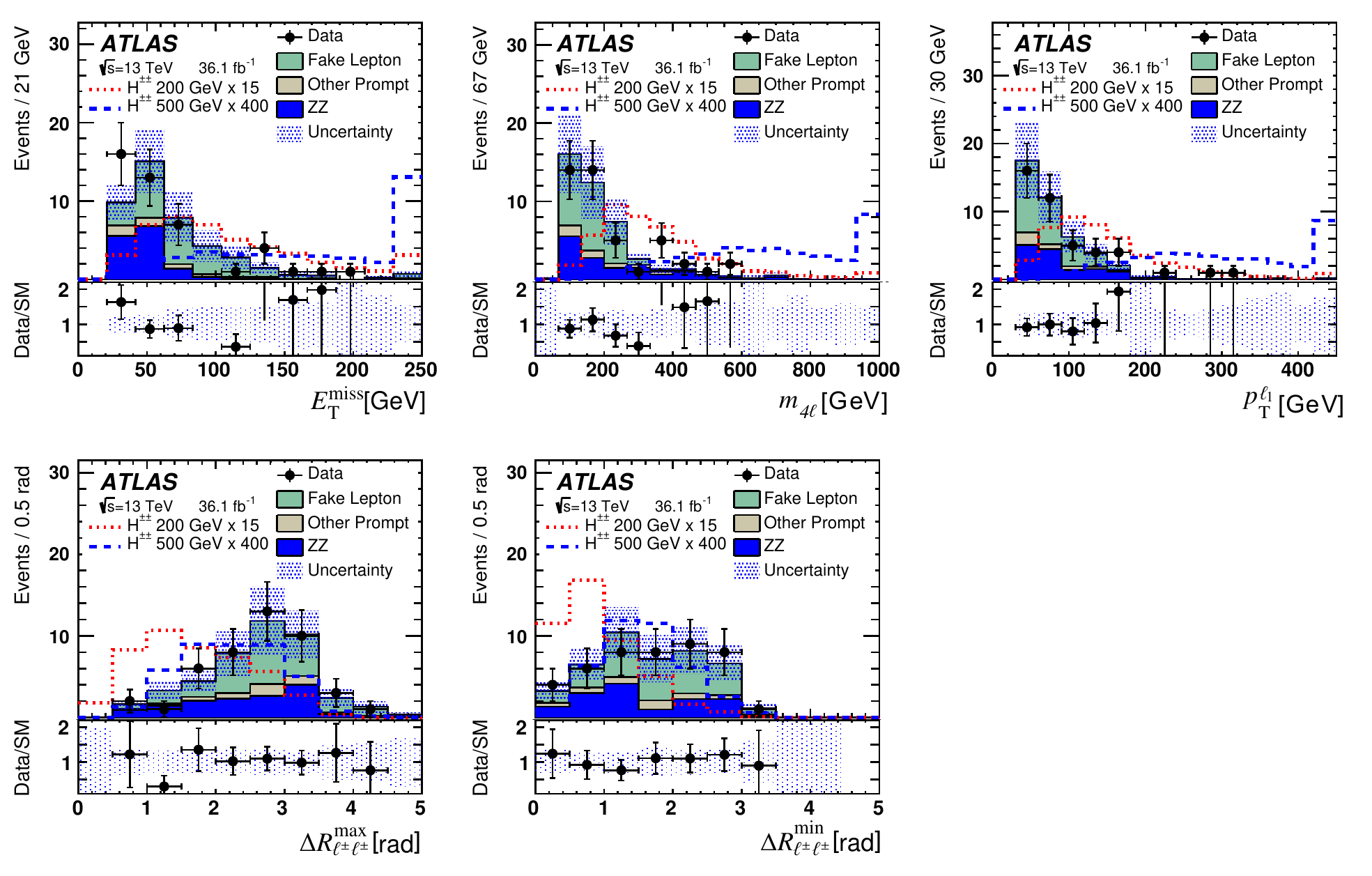}
        \caption{ Distribution of variables used for the signal region optimisation of the \llll channel (a detailed description can be found  in the caption of Figure~\ref{fig,2L_PRES}).}
        \label{fig,4L_PRES}
\end{figure}

The strategy used to extract the signal is based on rectangular cut optimisation using the TMVA package~\cite{Hocker:2007ht}. For each $m_{H^{\pm\pm}}$ hypothesis, six signal regions are defined using the following lepton flavour content: in the \ll channel, three signal regions are optimised separately for $ee$, $e\mu$ and $\mu\mu$ channels; in the \lll channel, the signal regions are optimised separately for events with no same-flavour opposite-sign lepton pairs (SFOS 0, for which the SM background is small) and for events with one or two such pairs (SFOS 1,2); the \llll channel is treated globally, with no further lepton flavour distinction. 
The selection criteria used to define the signal regions are shown in Table~\ref{Tab,cuts}. The optimisation is performed as a function of the $H^{\pm\pm}$ mass for $m_{H^{\pm\pm}}=200, 300, 400$ and $500$~\GeV, and seeks the best expected signal significance. The last optimisation point is applied to $m_{H^{\pm\pm}} \ge 600$~\GeV as well, since the signal discrimination power does not vary significantly in this regime. 

In order to verify the background estimate reliability for the signal region, three further checks were performed: the optimised cuts were applied individually, the cuts were applied successively, or each cut was inverted while the other cuts were applied. The agreement between data and prediction remains adequate for all those cases.

\begin{table}[htbp]
\footnotesize
 \begin{center}
 \caption{\label{Tab,cuts} The selection criteria used to define the signal regions. The variables are described in Section~\ref{sec,var}.}
   \begin{tabular}{l|c|c|c|c|c|c}
\hline
    & \multicolumn{3}{c|}{ \ll }    & \multicolumn{2}{c|}{ \lll }    & \multicolumn{1}{c}{ \llll } \\ 
    \cline{1-6}
Selection criteria                 & $e^\pm e^\pm$  & $e^\pm\mu^\pm$ & $\mu^\pm\mu^\pm$ & SFOS 0 & SFOS 1,2  & \\
\hline
  & \multicolumn{6}{c}{  $m_{H^{\pm\pm}}=200$~\GeV} \\
\hline

$\MET$[\GeV]                        & $>100$     &$>100$      & $>100$     &  $ > 45$            &  $> 45$           & $>60$  \\
$m_{x\ell}$[\GeV]                   &$[25,130]$  & $[15,150]$ & $[35,150]$ & $ > 160 $           &   $ > 170 $       & $>230$ \\
$\Delta R_{\ell^{\pm}\ell^{\pm}} $[rad.] &$<0.8$      & $<1.8$     & $<0.9$     &   $ [0.15 ,1.57] $  &   $ [0.00,1.52] $ &  \\
$\Delta\phi(\ell\ell,\MET) $[rad.]       &$<1.1$      & $<1.3$     &$<1.3$      &                     &                   & \\
$S$[rad.]                              & $<0.3$     &$<0.3$      &$<0.2 $     &                     &                   & \\
$m_\mathrm{jets}$[\GeV]                &$[140,770]$ & $[95,330]$ &$[95,640]$  &                     &                   & \\
 $\Delta R_{\ell-\mathrm{jet}} $[rad.]               &            &            &            &   $ [0.08 ,1.88] $  &  $ [0.07 ,1.31] $ &              \\
$ \pt^{\text{leading jet}}$[\GeV] &            &            &            & $> 80 $             &   $ > 55$         &     \\
$ \pt^{\ell_1}$[\GeV]               &            &            &            &                     &                   &   $ > 65$ \\
$ \Delta R^\text{min}_{\ell^{\pm}\ell^{\pm}}$[rad.] &      &            &            &                     &                   &    $ [0.16,1.21] $ \\  
$ \Delta R^\text{max}_{\ell^{\pm}\ell^{\pm}}$[rad.] &      &            &            &                     &                   &    $ [0.27,2.03] $ \\  

\hline
  & \multicolumn{6}{c}{  $m_{H^{\pm\pm}}=300$~\GeV} \\
\hline

$\MET$[\GeV]                        & $>200$     &$>200$      & $>200$     &  $ > 65$            &  $> 55$           & $>60$  \\
$m_{x\ell}$[\GeV]                   &$[105,340]$ & $[80,320]$ & $[80,320]$ & $ > 170 $           &   $ > 210 $       & $>270$ \\
$\Delta R_{\ell^{\pm}\ell^{\pm}} $[rad.] &$<1.4$      & $<1.8$     & $<1.8$     &   $ [0.18 ,2.23] $  &   $ [0.08,2.23] $ &  \\
$\Delta\phi(\ell\ell,\MET) $[rad.]       &$<2.1$      & $<2.4$     &$<2.4$      &                     &                   & \\
$S$[rad.]                              & $<0.4$     &$<0.4$      &$<0.4 $     &                     &                   & \\
$m_\mathrm{jets}$[\GeV]                    &$[180,770]$ & $[130,640]$&$[130,640]$ &                     &                   & \\
 $\Delta R_{\ell j} $[rad.]              &            &            &            &   $ [0.27 ,2.37] $  &  $ [0.21 ,2.08] $ &              \\
$ \pt^{\text{leading jet}}$[\GeV] &            &            &            & $> 95 $             &   $ > 80$         &     \\
$ \pt^{\ell_1}$[\GeV]               &            &            &            &                     &                   &   $ > 45$ \\
$ \Delta R^\text{min}_{\ell^{\pm}\ell^{\pm}}$[rad.] &      &            &            &                     &                   &    $ [0.09,1.97] $ \\  
$ \Delta R^\text{max}_{\ell^{\pm}\ell^{\pm}}$[rad.] &      &            &            &                     &                   &    $ [0.44,2.68] $ \\  

\hline
  & \multicolumn{6}{c}{  $m_{H^{\pm\pm}}=400$~\GeV} \\
\hline

$\MET$[\GeV]                        & $>200$     &$>200$      & $>200$     &  $ > 65$            &  $> 85$           & $>60$  \\
$m_{x\ell}$[\GeV]                   &$[105,340]$ & $[80,350]$ & $[80,350]$ & $ > 230 $           &   $ > 250 $       & $>270$ \\
$\Delta R_{\ell^{\pm}\ell^{\pm}} $[rad.] &$<2.2$      & $<1.8$     & $<1.8$     &   $ [0.22 ,2.39] $  &   $ [0.29,2.69] $ &  \\
$\Delta\phi(\ell\ell,\MET) $[rad.]       &$<2.4$      & $<2.4$     &$<2.4$      &                     &                   & \\
$S$[rad.]                              & $<0.6$     &$<0.6$      &$<0.5 $     &                     &                   & \\
$m_\mathrm{jets}$[\GeV]                    &$[280,1200]$&$[220,1200]$&$[220,1200]$&                     &                   & \\
 $\Delta R_{\ell j} $[rad.]              &            &            &            &   $ [0.30 ,2.59] $  &  $ [0.31 ,2.30] $ &              \\
$ \pt^{\text{leading jet}}$[\GeV] &            &            &            & $> 120 $            &   $ > 100$        &     \\
$ \pt^{\ell_1}$[\GeV]               &            &            &            &                     &                   &   $ > 110$ \\
$ \Delta R^\text{min}_{\ell^{\pm}\ell^{\pm}}$ [rad.]&      &            &            &                     &                   &   $ [0.39,2.22] $\\  
$ \Delta R^\text{max}_{\ell^{\pm}\ell^{\pm}}$[rad.] &      &            &            &                     &                   &     $ [0.55,2.90] $\\

\hline
  & \multicolumn{6}{c}{  $m_{H^{\pm\pm}}=500$--$700$~\GeV} \\
\hline

$\MET$[\GeV]                        & $>250$     &$>250$      & $>250$     &  $ > 120$           &  $> 100$          & $>60$  \\
$m_{x\ell}$[\GeV]                   &$[105,730]$ & $[110,440]$ & $[110,440]$& $ > 230 $           &   $ > 300 $       & $>370$ \\
$\Delta R_{\ell^{\pm}\ell^{\pm}} $[rad.] &$<2.6$      & $<2.2$     & $<2.2$     &   $ [0.39 ,3.11] $  &   $ [0.29,2.85] $ &  \\
$\Delta\phi(\ell\ell,\MET) $[rad.]      &$<2.6$      & $<2.4$     &$<2.4$      &                     &                   & \\
$S$[rad.]                              & $<1.1$     &$<1.1$      &$<1.1 $     &                     &                   & \\
$m_\mathrm{jets}$[\GeV]                    &$>440$      & $>470$     &$>470$      &                     &                   & \\
 $\Delta R_{\ell j} $[rad.]              &            &            &            &   $ [0.60 ,2.68] $  &  $ [0.31 ,2.53] $ &              \\
$ \pt^{\text{leading jet}}$[\GeV] &            &            &            & $> 130 $            &   $ > 130$        &     \\
$ \pt^{\ell_1}$[\GeV]               &            &            &            &                     &                   &   $ > 160$ \\
$ \Delta R^\text{min}_{\ell^{\pm}\ell^{\pm}}$ [rad.]&      &            &            &                     &                   &    $ [0.53,3.24] $ \\  
$ \Delta R^\text{max}_{\ell^{\pm}\ell^{\pm}}$[rad.] &      &            &            &                     &                   &    $ [0.59,2.94] $ \\  

\hline
   \end{tabular}
   
 \end{center}
\end{table}

\section{Systematic uncertainties}
\label{sec:Systematics}

The theoretical uncertainties associated with the signal prediction originate from the PDFs, the matrix element calculation and the parton shower simulation. The uncertainties related to PDFs are evaluated using the Hessian method provided in LHAPDF6~\cite{Buckley2015} 
and are found to be in the range from $2.5$\% to $4.5$\%. 
The uncertainty of the parton shower simulation is assessed by comparing PYTHIA (with A14 tune) and \Herwigpp (with UEEE5 tune~\cite{HerwigPPUE}), and is found to be $2.4$\%, $1.7$\%, and $3.8$\% for the \ll, \lll, and \llll channels, respectively. The higher-order corrections are assumed to induce an additional $15$\% uncertainty in the cross-section calculations~\cite{Muhlleitner:2003me}. Combining those uncertainties in quadrature, an overall uncertainty of $17$\% is obtained for the signal normalisation. 

The theoretical uncertainties associated with the largest SM backgrounds, $VV$~\cite{ATL-PHYS-PUB-2017-005}
 (including same-sign $WWqq$ and $WZ$ processes) and
 $t\overline{t}V$~\cite{ATL-PHYS-PUB-2017-007}, are estimated using dedicated MC
 samples, where the factorisation and renormalisation scales are varied independently by factors of $2$ and $0.5$ and the parton shower parameters are varied within the given model uncertainties. The theoretical uncertainties
 obtained are $24$\% for $VV$ and $17$\% for $t\overline{t}V$. The uncertainties related to PDFs are found to be negligible. The uncertainty associated
 with the $V\gamma$ contributions is taken to be 25\%, as indicated by dedicated studies using converted photons. 
The uncertainties related to the $VVV$ and $tZ$ process predictions are taken from the respective
 inclusive cross-section measurements~\cite{ATLAS-WWW,TOPQ-2016-14}.
For other rare backgrounds which have no dedicated measurements yet ($tt\bar{t}$, $t\bar{t}W^+W^-$), uncertainties of $50$\% are 
assumed and are found to have a negligible impact on the sensitivity. 

The experimental uncertainties arise from the accuracy of the detector simulation and from the uncertainties associated
with the data-driven methods that are used to estimate the instrumental backgrounds. These uncertainties originate from the following sources: 
\begin{itemize}
\item 
The uncertainties related to event reconstruction include the lepton~\cite{ATL-PHYS-PUB-2016-015, PERF-2015-10} and the jet~\cite{PERF-2016-04} energy 
scales and resolutions and the uncertainties in the reconstruction of \MET~\cite{PERF-2016-07}. 
The impact of this type of uncertainty on the signal and background yields 
is in the range $3\%$--$8\%$ and $10\%$--$30\%$, respectively. 
\item The uncertainties related to the efficiencies of electron~\cite{ATLAS-CONF-2016-024} and muon~\cite{PERF-2015-10} 
reconstruction and identification, including the uncertainties in the trigger efficiency are estimated in dedicated studies. 
The impact of this type of uncertainty on the 
signal and background yields is found to be in the range $4\%$--$6\%$ and $2\%$--$5\%$, respectively. The uncertainties related to the $b$-jet identification algorithms, used in the analysis to veto events containing $b$-jets, are found to be negligible.
\item The uncertainties originating from data-taking conditions include the luminosity measurement and the pile-up simulation procedure. The 
uncertainty of the integrated luminosity is $2.1\%$, determined using a methodology similar to that detailed in Ref.~\cite{DAPR-2013-01}. 
The uncertainty from the pile-up simulation is about $5\%$.
\item The uncertainties related to the background contributions from electron charge misidentification are $22\%$ to $28\%$ for the \ll $ee$ and $e\mu$ channels. The uncertainties of the fake-lepton contributions range from $50\%$ to $250$\%, and mainly originate from the fake factors (\ll and \lll channels) and the scale factors for MC simulation (\llll channel) described in Section~\ref{sec:fakeLeptons} and the statistics of the control samples. The uncertainties exceed $100\%$ in some cases due to the subtraction of the prompt-lepton contributions in the fake-lepton control regions.   
\end{itemize}

The theoretical and experimental systematic uncertainties described above are
assumed to be correlated amongst the various signal regions in the interpretation of the final results. 
Overall, the sensitivity of the search is dominated by the statistical uncertainty of the event yield in the signal regions.

\section{Results}
\label{sec:Results}

The expected and observed event yields in the signal regions are shown in Figure~\ref{fig:SByields} and Table~\ref{Tab:EventYield}. For a $H^{\pm\pm}$ mass of 200~\GeV, substantial signal yield is expected in all channels, and the analysis sensitivity is found to be comparable across the \ll, \lll 
 and \llll channels. No significant excess has been observed. Table~\ref{Tab:EventYield} also includes the overall signal acceptance $A$, defined as the number of selected events selected in a given channel divided by the total number of \hppp events and representing the signal reduction due to phase space acceptance, branching ratio and detector efficiency.
 
 The statistical analysis of the results is based on a likelihood ratio test~\cite{CLREF} using the CL$_\mathrm{s}$ method~\cite{cls}. The parameter of interest is the signal strength,  defined as the cross-section of the hypothetical contribution from physics beyond the SM in units of the cross-section of the benchmark model. 
 The likelihood function is constructed from Poisson probability distributions of counting experiments for each of the six channels in each signal region. The systematic uncertainties are treated as  nuisance parameters implemented in the likelihood functions with Gaussian constraints.
 
 The expected and observed upper limits of the \hppd cross-section at 95\% confidence level (CL), obtained from the combination of \ll, \lll 
and \llll channels for the six \hpp mass hypotheses are shown in Figure~\ref{fig:ExpectedLimits_comb}. Assuming a linear 
interpolation of the sensitivity between neighbouring mass hypotheses, and the cross-section of the benchmark model, the observed (expected) lower limit on the mass of the \hpp boson is 220~\GeV (250~\GeV) at 95\% CL.

\begin{table}[htbp]
\footnotesize
\renewcommand{\arraystretch}{1.1}
 \begin{center}
  \caption{\label{Tab:EventYield} Event yields in the signal regions of corresponding targeted masses of $H^{\pm\pm}$. The signal yield is for the corresponding mass point and is normalised to the luminosity of 36.1~\ifb.  The dominant background from prompt-lepton sources is from the $WZ$ process in the \ll channel. For the \lll and \llll channels, the dominant background from prompt-lepton sources is from $WZ$ and $t\overline{t}V$ processes. The overall signal acceptance $A$ and the upper limit of extra contribution to each signal region at 95\% confidence level $n_{95}$ are also presented. The data and SM prediction yields obtained for $m_{H^{\pm\pm}}=500$~\GeV are also valid for $m_{H^{\pm\pm}}=600$ and $700$~\GeV.}  
   \begin{tabular}{l|c|c|c|c|c|c}
\hline
    & \multicolumn{3}{c|}{ \ll }    & \multicolumn{2}{c|}{ \lll }    & \multicolumn{1}{c}{ \llll } \\ 
    \cline{1-6}
Subchannel               & $e^\pm e^\pm$  & $e^\pm\mu^\pm$ & $\mu^\pm\mu^\pm$ & SFOS 0 & SFOS 1,2  & \\
\hline
   \multicolumn{7}{c}{  $m_{H^{\pm\pm}}=200$~\GeV} \\
\hline
Prompt lepton           & $0.5 \pm 0.2$  & $0.3\pm0.2$     & $1.3\pm0.6$  & $0.3\pm0.1$  & $1.4 \pm0.5$   & $0.07 \pm 0.03$    \\
QMisID           & $0.6 \pm 0.2 $ & $0.4\pm0.1$     & $-$            & $-$            & $-$             & $-$   \\
Fake lepton            & $1 \pm 1 $ & $     < 0.4$    & $0.4\pm0.3$  & $0.2\pm0.1$  & $0.2\pm0.1$   & $0.03 \pm 0.02$  \\
\hline 
Total background & $2\pm1$    &$0.6 \pm0.3$     &$1.7\pm0.7$   & $0.5\pm0.1$  & $1.7\pm0.6$   & $0.11\pm0.05$  \\
\hline
Signal           & $1.1 \pm 0.2$  &$2.3\pm0.4$      &$2.4\pm0.4$   &$1.8\pm 0.3$  & $5.0\pm0.9$   & $1.1\pm0.2$   \\
$A$ [\%]         & $0.037$  &$0.080$      &$0.082$   &$0.061$  & $0.17$   & $0.038$    \\
$n_{95}$         & $12.3$  &$7.1$      &$7.5$   &$4.1$  & $7.7$   & $3.8$    \\

\hline
Data & 3 & 2  & 2 & 1& 2& 0\\
\hline 
 \multicolumn{7}{c}{  $m_{H^{\pm\pm}}=300$~\GeV} \\
\hline
Prompt lepton           & $0.1\pm0.1$   & $0.9\pm0.4$ & $0.02\pm0.02$   & $0.4\pm0.1$ &  $4\pm1$& $0.3 \pm 0.1$  \\
QMisID           & $0.1\pm0.1$   & $0.07\pm0.04$ & $-$             & $-$           &  $-$          & $-$   \\
Fake lepton            & $0.4\pm0.5$   & $<0.2$       & $<0.4$         & $0.3\pm0.2$ & $0.8\pm 0.4$& $0.2 \pm 0.2$ \\
\hline 
Total background & $0.7\pm0.5$   &$1.0\pm0.5$  &$0.02\pm0.02$    & $0.8\pm0.2$ & $5\pm2$ & $0.5\pm0.2$  \\
\hline
Signal           &$0.16\pm0.03$    & $0.6\pm0.1$ &$0.29\pm0.05$    & $0.6\pm0.1$ & $1.8\pm0.3$ & $0.43\pm0.08$\\
$A$ [\%]         & $0.027$  &$0.10$      &$0.049$   &$0.11$  & $0.30$   & $0.071$    \\
$n_{95}$                            & $4.0$  &$9.6$      &$3.0$   &$3.1$  & $22.7$   & $3.8$    \\

\hline
Data & 0 &3 & 0 & 0 & 11 & 0 \\
\hline 
 \multicolumn{7}{c}{  $m_{H^{\pm\pm}}=400$~\GeV} \\
\hline
Prompt lepton          & $0.7\pm0.3$  & $1.0\pm0.4$    &$0.2\pm0.1$          & $0.3\pm0.1$ & $4\pm1$ &  $ 0.3 \pm 0.1 $ \\
QMisID           & $0.3\pm0.1$  & $0.2\pm0.1$    &$-$                    & $-$           & $-$           &   $-$   \\
Fake lepton            & $0.4\pm0.5$  & $<0.3$          &$<0.4$                & $0.3\pm0.2$ & $0.2\pm0.1$ &  $0.05\pm 0.04$ \\
\hline 
Total background & $1.4\pm0.6$  &$1.2\pm0.5$     &$0.3\pm0.1$          & $0.6\pm0.2$ & $4\pm1$ &     $0.4\pm 0.1$      \\
\hline
Signal           & $0.20\pm0.04$  &$0.38\pm0.07$     &$0.19\pm0.03$          & $0.23\pm0.04$ & $0.6\pm0.1$ & $0.17\pm0.03 $\\
$A$ [\%]         & $0.11$  &$0.21$      &$0.11$   &$0.13$  & $0.36$   & $0.092$    \\
$n_{95}$        & $10.4$  &$18.3$      &$6.4$   &$3.1$  & $10.4$   & $4.3$    \\

\hline
Data & $2$ &$6$ & $1$ & 0 & 4 & 1\\
\hline 
 \multicolumn{7}{c}{  $m_{H^{\pm\pm}}=500$~\GeV} \\
\hline
Prompt lepton          & $1.0\pm0.4$  & $0.7\pm0.3$ & $0.3\pm0.2$  & $0.4\pm0.1$ &   $3\pm1$  & $0.2\pm0.1$ \\
QMisID           & $0.3\pm0.1$  & $0.2\pm0.1$ & $-$            & $-$           &   $-$            & $-$ \\
Fake lepton            & $0.2\pm0.5$  & $0.3\pm0.5$ & $ <0.4$         & $ 0.11\pm0.06$&   $0.10\pm 0.05$ & $0.2\pm0.2$  \\
\hline 
Total background & $1.6\pm0.6$  & $1.2\pm0.6$ & $0.3\pm0.2$  & $0.5\pm 0.1$&   $3.0\pm0.8$  & $0.4\pm0.2$ \\
\hline
Signal           & $0.10\pm0.02$  & $0.16\pm0.03$ & $0.07\pm0.01$  &$0.09\pm0.02$ &    $0.24\pm0.04$  & $0.06\pm0.01$\\
$A$ [\%]         & $0.16$  &$0.25$      &$0.11$   &$0.14$  & $0.37$   & $0.098$    \\
$A$ [\%]   $m_{H^{\pm\pm}}=600$~\GeV      & $0.22$  &$0.36$      &$0.16$   &$0.17$  & $0.44$   & $0.11$    \\
$A$ [\%]    $m_{H^{\pm\pm}}=700$~\GeV       & $0.26$  &$0.38$      &$0.17$   &$0.19$  & $0.48$   & $0.12$    \\
$n_{95}$        & $8.6$  &$12.7$      &$3.8$   &$3.0$  & $7.9$   & $4.9$    \\

\hline
Data & $4$ &$3$ & $0$ & 0& 2 & 3\\
\hline 

   \end{tabular}
 \end{center}
\end{table}

\begin{figure}[htp]        
 \centering
 \subfloat[ 200~\GeV]{\includegraphics[width=0.45\textwidth]{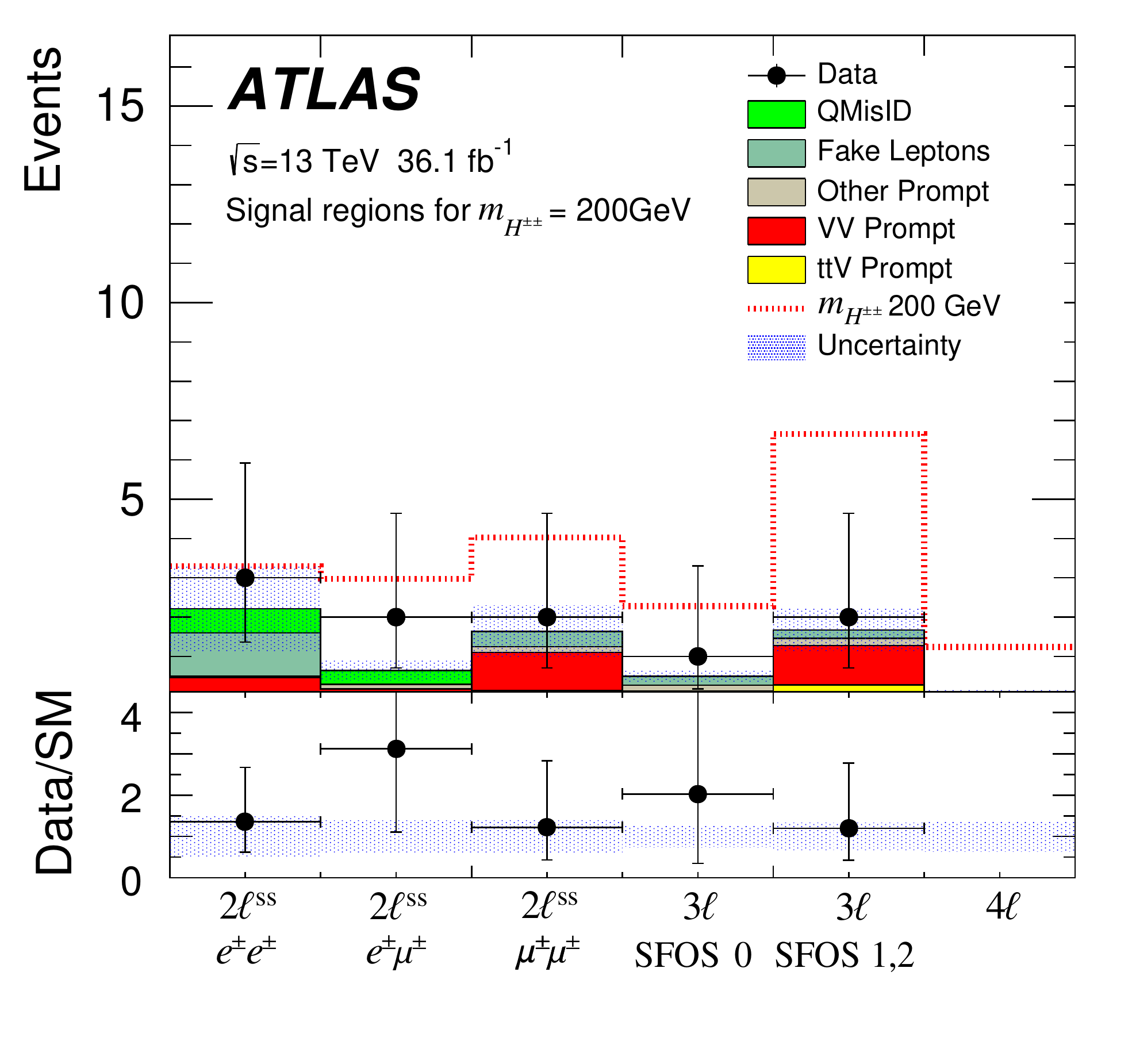}}   
        \qquad
        \subfloat[ 300~\GeV]{\includegraphics[width=0.45\textwidth]{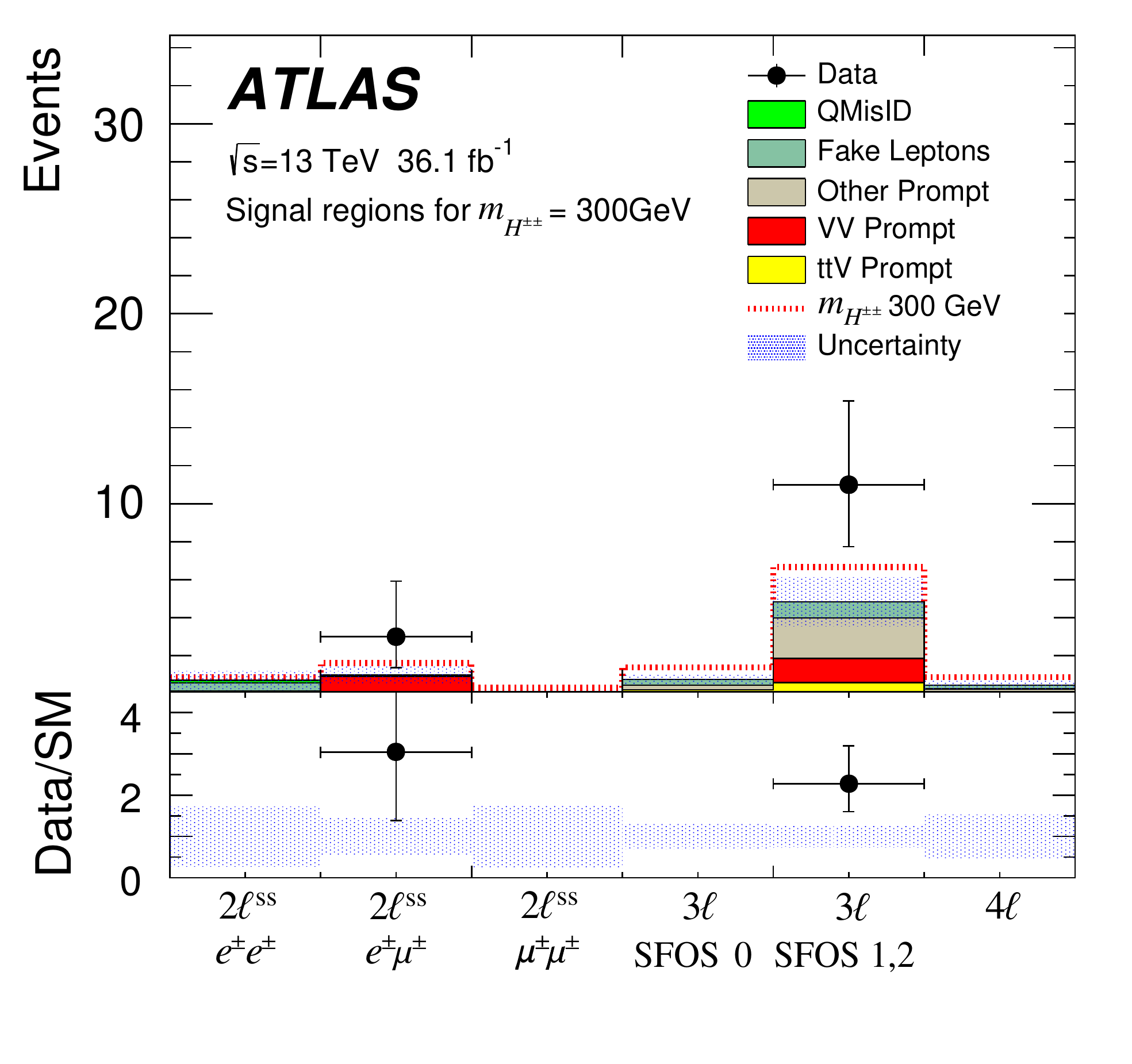}}
        \qquad
  \subfloat[ 400~\GeV]{\includegraphics[width=0.45\textwidth]{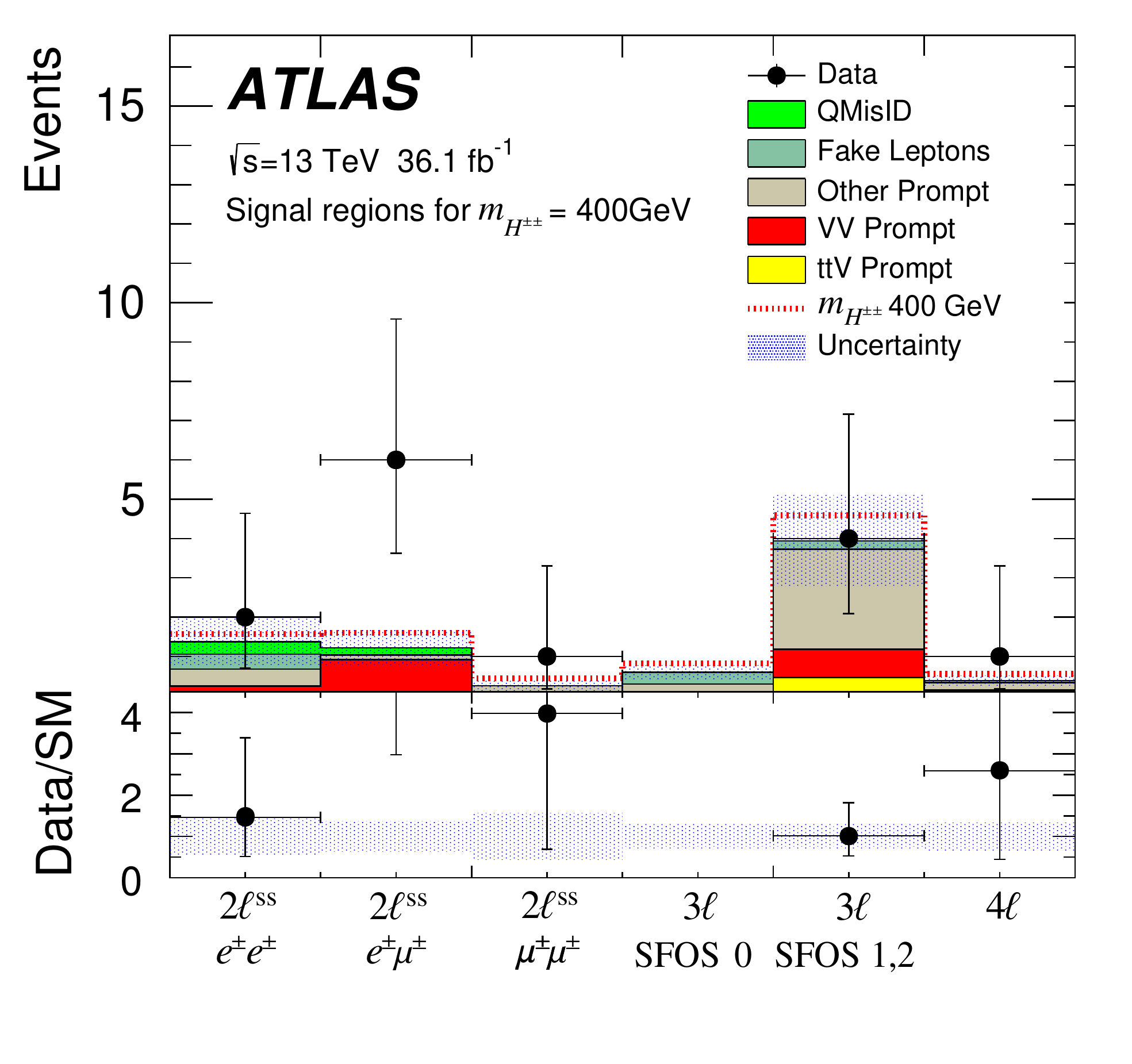}}   
        \qquad
        \subfloat[ 500~\GeV]{\includegraphics[width=0.45\textwidth]{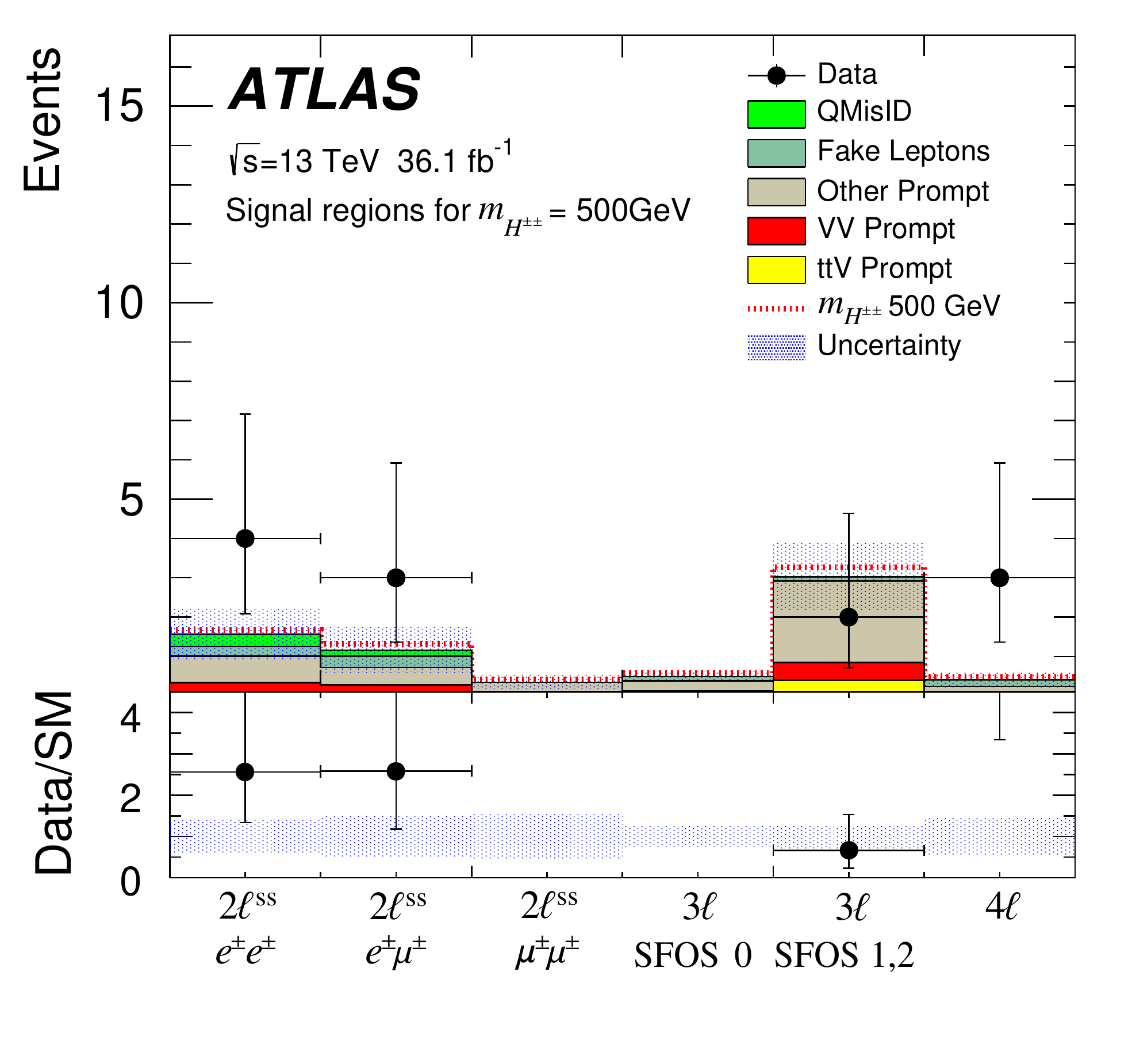}}
          \qquad

 \caption{Event yields in the signal regions optimised for the  $m_{H^{\pm\pm}}$= $200$, $300$, $400$ and $500$~\GeV searches. The bottom panel shows the ratio of the data to the total background prediction, where the band illustrates the total uncertainty of the SM background. The error bars attributed to data are estimated assuming a Poisson distribution with the average equal to the respective yields.  The signal prediction is represented as a dotted histogram, stacked on the SM background. }
       \label{fig:SByields}
\end{figure}

\begin{figure}
 \includegraphics[width=.80\linewidth]{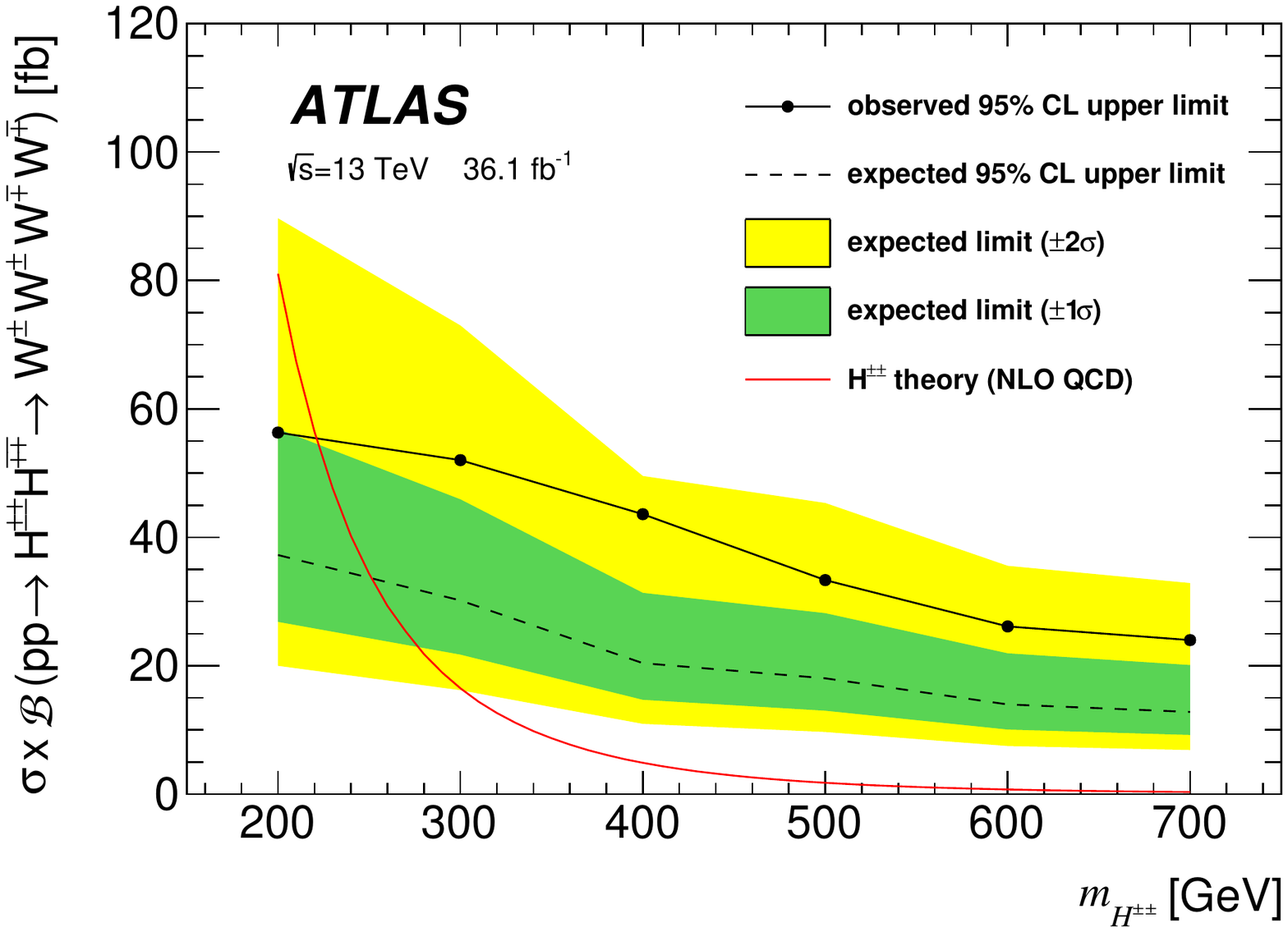}
 \caption{Observed and expected upper limits for \hppp cross-section times branching fraction at $95\%$ CL obtained from the combination of \ll, \lll and \llll channels. The region above the observed limit is excluded by the measurement. The bands represent the expected exclusion curves within one and two standard deviations. The theoretical prediction~\cite{Arhrib:2011uy} including the NLO QCD corrections~\cite{Muhlleitner:2003me}  is also shown and is excluded for $\mhh$~$<$~220~\GeV. }
\label{fig:ExpectedLimits_comb}
\end{figure}

\FloatBarrier
\section{Conclusion}
\label{sec:Conclusions}
A search for the pair production of doubly charged Higgs scalar bosons with subsequent 
decays into $W$ bosons is performed in proton--proton collisions at a centre-of-mass 
energy of $13$~\TeV. The data sample was
collected by the ATLAS experiment at the LHC and corresponds to an integrated luminosity of $36.1$~\ifb. The search for the $H^{\pm\pm} \to W^{\pm} W^{\pm}$ decay mode, not considered in previous analyses at colliders, is motivated by a model with an extended scalar sector that includes a triplet in addition to the Standard Model scalar doublet. The analysis proceeds through 
the selection of multi-lepton events in three channels (a pair of same-sign leptons, three leptons and four leptons) 
with missing transverse momentum and jets. 
The signal region is optimised as a function of the  $H^{\pm\pm}$ mass.
The data are found to be in good agreement with the Standard Model predictions for all channels investigated.
Combining those channels, the model considered is excluded at 95\% confidence level for \hpp boson masses between $200$ and $220$~\GeV.

\section*{Acknowledgements}

We thank CERN for the very successful operation of the LHC, as well as the
support staff from our institutions without whom ATLAS could not be
operated efficiently.

We acknowledge the support of ANPCyT, Argentina; YerPhI, Armenia; ARC, Australia; BMWFW and FWF, Austria; ANAS, Azerbaijan; SSTC, Belarus; CNPq and FAPESP, Brazil; NSERC, NRC and CFI, Canada; CERN; CONICYT, Chile; CAS, MOST and NSFC, China; COLCIENCIAS, Colombia; MSMT CR, MPO CR and VSC CR, Czech Republic; DNRF and DNSRC, Denmark; IN2P3-CNRS, CEA-DRF/IRFU, France; SRNSFG, Georgia; BMBF, HGF, and MPG, Germany; GSRT, Greece; RGC, Hong Kong SAR, China; ISF and Benoziyo Center, Israel; INFN, Italy; MEXT and JSPS, Japan; CNRST, Morocco; NWO, Netherlands; RCN, Norway; MNiSW and NCN, Poland; FCT, Portugal; MNE/IFA, Romania; MES of Russia and NRC KI, Russian Federation; JINR; MESTD, Serbia; MSSR, Slovakia; ARRS and MIZ\v{S}, Slovenia; DST/NRF, South Africa; MINECO, Spain; SRC and Wallenberg Foundation, Sweden; SERI, SNSF and Cantons of Bern and Geneva, Switzerland; MOST, Taiwan; TAEK, Turkey; STFC, United Kingdom; DOE and NSF, United States of America. In addition, individual groups and members have received support from BCKDF, CANARIE, CRC and Compute Canada, Canada; COST, ERC, ERDF, Horizon 2020, and Marie Sk{\l}odowska-Curie Actions, European Union; Investissements d' Avenir Labex and Idex, ANR, France; DFG and AvH Foundation, Germany; Herakleitos, Thales and Aristeia programmes co-financed by EU-ESF and the Greek NSRF, Greece; BSF-NSF and GIF, Israel; CERCA Programme Generalitat de Catalunya, Spain; The Royal Society and Leverhulme Trust, United Kingdom. 

The crucial computing support from all WLCG partners is acknowledged gratefully, in particular from CERN, the ATLAS Tier-1 facilities at TRIUMF (Canada), NDGF (Denmark, Norway, Sweden), CC-IN2P3 (France), KIT/GridKA (Germany), INFN-CNAF (Italy), NL-T1 (Netherlands), PIC (Spain), ASGC (Taiwan), RAL (UK) and BNL (USA), the Tier-2 facilities worldwide and large non-WLCG resource providers. Major contributors of computing resources are listed in Ref.~\cite{ATL-GEN-PUB-2016-002}.

\clearpage
\printbibliography

\clearpage 
 
\begin{flushleft}
{\Large The ATLAS Collaboration}

\bigskip

M.~Aaboud$^\textrm{\scriptsize 34d}$,    
G.~Aad$^\textrm{\scriptsize 99}$,    
B.~Abbott$^\textrm{\scriptsize 124}$,    
O.~Abdinov$^\textrm{\scriptsize 13,*}$,    
B.~Abeloos$^\textrm{\scriptsize 128}$,    
D.K.~Abhayasinghe$^\textrm{\scriptsize 91}$,    
S.H.~Abidi$^\textrm{\scriptsize 164}$,    
O.S.~AbouZeid$^\textrm{\scriptsize 39}$,    
N.L.~Abraham$^\textrm{\scriptsize 153}$,    
H.~Abramowicz$^\textrm{\scriptsize 158}$
% ...
\bigskip
\\

% ...
$^{13}$Institute of Physics, Azerbaijan Academy of Sciences, Baku; Azerbaijan.\\
% ...
$^{34}$$^{(a)}$Facult\'e des Sciences Ain Chock, R\'eseau Universitaire de Physique des Hautes Energies - Universit\'e Hassan II, Casablanca;$^{(b)}$Centre National de l'Energie des Sciences Techniques Nucleaires (CNESTEN), Rabat;$^{(c)}$Facult\'e des Sciences Semlalia, Universit\'e Cadi Ayyad, LPHEA-Marrakech;$^{(d)}$Facult\'e des Sciences, Universit\'e Mohamed Premier and LPTPM, Oujda;$^{(e)}$Facult\'e des sciences, Universit\'e Mohammed V, Rabat; Morocco.\\
% ...
$^{39}$Niels Bohr Institute, University of Copenhagen, Copenhagen; Denmark.\\
% ...
$^{91}$Department of Physics, Royal Holloway University of London, Egham; United Kingdom.\\
% ...
$^{99}$CPPM, Aix-Marseille Universit\'e, CNRS/IN2P3, Marseille; France.\\
% ...
$^{124}$Homer L. Dodge Department of Physics and Astronomy, University of Oklahoma, Norman OK; United States of America.\\
% ...
$^{128}$LAL, Universit\'e Paris-Sud, CNRS/IN2P3, Universit\'e Paris-Saclay, Orsay; France.\\
% ...
$^{153}$Department of Physics and Astronomy, University of Sussex, Brighton; United Kingdom.\\
% ...
$^{158}$Raymond and Beverly Sackler School of Physics and Astronomy, Tel Aviv University, Tel Aviv; Israel.\\
% ...
$^{164}$Department of Physics, University of Toronto, Toronto ON; Canada.\\
% ...
$^{a}$ Also at Borough of Manhattan Community College, City University of New York, NY; United States of America.\\
$^{b}$ Also at Centre for High Performance Computing, CSIR Campus, Rosebank, Cape Town; South Africa.\\
$^{c}$ Also at CERN, Geneva; Switzerland.\\
$^{d}$ Also at CPPM, Aix-Marseille Universit\'e, CNRS/IN2P3, Marseille; France.\\
$^{e}$ Also at D\'epartement de Physique Nucl\'eaire et Corpusculaire, Universit\'e de Gen\`eve, Gen\`eve; Switzerland.\\
% ...
$^{*}$ Deceased

\end{flushleft}


\end{document}